\documentclass[botnum, fleqn]{unmeethesis}

\usepackage{subcaption}
\usepackage{verbatim}
\usepackage{graphicx}
\usepackage{xurl}
\usepackage{multirow}
\usepackage{tipa}
\usepackage[T1]{fontenc}
\usepackage{xcolor}

\usepackage{listings}
\usepackage{color}

\definecolor{codegreen}{rgb}{0,0.6,0}
\definecolor{codegray}{rgb}{0.5,0.5,0.5}
\definecolor{codepurple}{rgb}{0.58,0,0.82}
\definecolor{backcolour}{rgb}{0.95,0.95,0.92}

\lstdefinestyle{mystyle}{
	backgroundcolor=\color{backcolour},   
	commentstyle=\color{codegreen},
	keywordstyle=\color{magenta},
	numberstyle=\tiny\color{codegray},
	stringstyle=\color{codepurple},
	basicstyle=\footnotesize,
	breakatwhitespace=false,         
	breaklines=true,                 
	captionpos=b,                    
	keepspaces=true,                 
	numbers=left,                    
	numbersep=5pt,                  
	showspaces=false,                
	showstringspaces=false,
	showtabs=false,                  
	tabsize=2
}
\DeclareUnicodeCharacter{029D}{\textctj}
\DeclareUnicodeCharacter{03B2}{\textbeta}
\DeclareUnicodeCharacter{0283}{\textesh}
\DeclareUnicodeCharacter{025B}{\textepsilon}
\DeclareUnicodeCharacter{027E}{\textfishhookr}
\DeclareUnicodeCharacter{02D0}{\textlengthmark}
\DeclareUnicodeCharacter{0272}{\textltailn}
\DeclareUnicodeCharacter{0259}{\textschwa}
\DeclareUnicodeCharacter{026A}{\textsci}
\DeclareUnicodeCharacter{0251}{\textscripta}
\DeclareUnicodeCharacter{0279}{\textturnr}
\DeclareUnicodeCharacter{028A}{\textupsilon}
\DeclareUnicodeCharacter{0254}{\textopeno}
\DeclareUnicodeCharacter{0263}{\textgamma}
\DeclareUnicodeCharacter{0261}{\textscriptg}
\DeclareUnicodeCharacter{025A}{\textrhookschwa}
\DeclareUnicodeCharacter{025C}{\textrevepsilon}
\DeclareUnicodeCharacter{028C}{\textturnv}
\DeclareUnicodeCharacter{0292}{\textyogh}
\DeclareUnicodeCharacter{03B8}{\texttheta}
\DeclareUnicodeCharacter{028E}{\textturny}
\DeclareUnicodeCharacter{00E6}{\ae}
\DeclareUnicodeCharacter{00F0}{\dh}
\DeclareUnicodeCharacter{00A0}{\ng}

\begin{document}

\frontmatter

\begin{flushleft}
    \underline{Mario Esparza}\\
    \emph{\small{Student Name}}\\[56pt]
    \underline{Department of Electrical and Computer Engineering}\\
    \emph{\small{Department}}\\[56pt]
    This thesis is approved and is acceptable in quality and form for publication.\\[12pt]
    \emph{Approved by the Thesis Committee:}\\[24pt]
    Dr. Mario Pattichis, Chair\\[18pt]
    Dr. Ramiro Jordan\\[18pt]
    Dr. Sylvia Celedón-Pattichis\\[18pt]
    Dr. Balasubramaniam Santhanam\\[24pt]
    
\end{flushleft}


\title{Spanish and English Phoneme Recognition by Training on Simulated Classroom Audio Recordings of Collaborative Learning Environments}

\author{Mario J. Esparza Perez}

\degreesubject{M.S., Computer Engineering}

\degree{Master of Science \\ Computer Engineering}

\documenttype{Thesis}

\previousdegrees{B.S., Computer Engineering, University of New Mexico, 2017}

\date{July, \thisyear}

\maketitle

\begin{dedication}
   To my family, and friends for their support,
   and encouragement. \\[3ex]
   ``Whatever your mind can conceive, your mind can achieve''
         -- Napoleon Hill
\end{dedication}

\begin{acknowledgments}
   \vspace{1.1in}
   First and foremost, I would like to express my sincere gratitude to my faculty advisor, Dr. Marios Pattichis. For being there every step of the way, commenting on results, recommending new approaches and patiently waiting when progress was slow. It would have certainly been a harder journey if it was not for his support.
    
    Similarly, I would like to thank my partner for her patience and support. If times were stressful or difficult she was always there. Additionally, I would like to thank my family and friends for their words of encouragement.

    Lastly, I would like to thank the colleagues from the lab, particularly Luis Sanchez as he worked closely with me throughout the whole journey.
    
    This material is based upon work supported by the National Science Foundation under the AOLME project (Grant No. 1613637), the AOLME Video Analysis project (Grant No. 1842220), and the ESTRELLA project (Grant No. 1949230). Any opinions or findings of this paper reflect the views of the author. They do not necessarily reflect the views of NSF.
\end{acknowledgments}

\maketitleabstract 

\begin{abstract}
Audio recordings of collaborative learning environments contain a constant presence of cross-talk and background noise. Dynamic speech recognition between Spanish and English is required in these environments. To eliminate the standard requirement of large-scale ground truth, the thesis develops a simulated dataset by transforming audio transcriptions into phonemes and using 3D speaker geometry and data augmentation to generate an acoustic simulation of Spanish and English speech.

The thesis develops a low-complexity neural network for recognizing Spanish and English phonemes (available at github.com/muelitas/keywordRec). When trained on 41 English phonemes, 0.099 PER is achieved on Speech Commands. When trained on 36 Spanish phonemes and tested on real recordings of collaborative learning environments, a 0.7208 LER is achieved. Slightly better than Google’s Speech-to-text 0.7272 LER, which used anywhere from 15 to 1,635 times more parameters and trained on 300 to 27,500 hours of real data as opposed to 13 hours of simulated audios.

\clearpage 
\end{abstract}

\tableofcontents
\listoffigures
\listoftables

\begin{glossary}{Longest  string}
\item[AOLME]
Advancing Out-of-School Learning in Mathematics
\item[ASR]
Automatic Speech Recognition
\item[CNN]
Convolutional Neural Network
\item[CTC]
Connectionist Temporal Classification \cite{CTC}. Useful when data is not segmented, i.e. when each character in the transcript is not aligned to the exact location of the audio recording. Given $L$ + 1 labels (including a $blank$ label) and an input sequence, CTC outputs a label score for each time step. This is performed by matching the input sequence with all possible permutations of label sequences and outputting a distribution of probabilities. Then, for each label sequence, the total probability is calculated by adding the probabilities of all different alignments that resulted in such sequence.
\item[E2E]
End-to-end
\item[GRU]
Gated Recurrent Unit
\item[ICA]
Independent Component Analysis
\item[KWS]
Keyword Spotting
\item[LER]
Label Error Rate
\item[LSTM]
Long Short-Term Memory
\item[PCA]
Principal Component Analysis
\item[PER]
Phoneme Error Rate
\item[RNN]
Recurrent Neural Network
\item[SOTA]
State-of-the-art
\item[WER]
Word Error Rate
\end{glossary}

\mainmatter

\chapter{Introduction}
\section{Automatic Speech Recognition}
ASR has been a topic of interest for decades \cite{old1} \cite{old2} \cite{old3}. Throughout time, researchers have had to implement different techniques to tackle the large list of areas in ASR. Among the most popular techniques, we have Hidden Markov Models, Gaussian Mixture Models, Deep Neural Networks, Convolutional Neural Networks, Recurrent Neural Networks, Transformers, and AutoEncoders. Additionally, one can also find implementations where the aforementioned techniques are used jointly. For example, \cite{transformerCNN} implements a transformer along with some convolutional layers. The network's architectures of \cite{ds2} and \cite{specaugment} are composed of multiple CNNs and RNNs. Graves et al. \cite{drnn} use a deep recurrent neural network.

These techniques have achieved breakthroughs in different areas of ASR. In keyword spotting, where one tries to only detect a set of keywords, \cite{res15} and \cite{kwt3} have accomplished SOTA results on the Google Speech Commands v2 Dataset \cite{sc}. In End-to-End Speech Recognition (recognition of all labels or words), \cite{wav2vec}, \cite{specaugment} and \cite{ds2} have shown great results on datasets like LibriSpeech \cite{ls}, TIMIT \cite{ti} and WSJ \cite{wsj}. When it comes to Noisy Speech Recognition (datasets like CHiME \cite{chime}), the ones that lead are \cite{chimesota}, \cite{kaldi} and \cite{ds2}.

Aside from this, multilingualism is yet another area of ASR. One involves multiple languages to be recognized, even though some of them might have limited resources (a small number of available recordings for example). On this end, M\"{u}ller et al. \cite{muller1} \cite{muller2} have shown great results on the Euronews Corpus \cite{euronews} by training a network capable of recognizing four different languages \cite{muller2}.

In addition to the aforementioned areas of ASR, there are two more problems worth mentioning, denoising and data augmentation. Removing background noise and increasing the size of limited-resources datasets are essential tasks in ASR. SpecAugment \cite{specaugment}, Random Erasing \cite{randomerase}, Pyroomacoustics \cite{pyroom} and Autoencoders \cite{ae3} \cite{ae4} are great examples of data augmentation techniques. The latter, however, is of particular interest to this thesis, since it can also perform denoising \cite{ae1} \cite{ae2}.
   
\section{Speech Recognition in Collaborative Learning Environments} \label{intro:aolme}
\begin{figure}[h!]
 \centering
 \includegraphics[width=0.8\columnwidth]{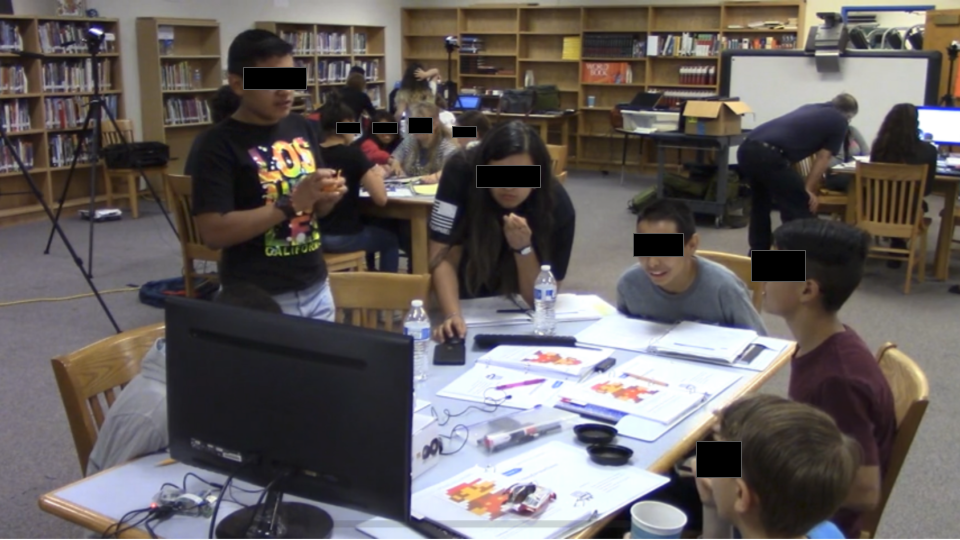}
    \caption[AOLME's classroom example]{Example of AOLME's classrooms and interactions. Notice the number of tables in the room as well as the number of students. Also, notice the single microphone placed in the middle of the table closest to the camera.}
    \label{fig:AOLMEclass}
\end{figure}

Despite the vast amount of research performed in ASR, we claim that speech recognition in bilingual collaborative learning environments remains challenging. The AOLME’s after-school program has offered multiple extracurricular activities to underrepresented middle school students in which STEM topics like python and image processing are taught. Figure \ref{fig:AOLMEclass} shows an example of a classroom setup in which these activities take place.

Each classroom includes six to eight tables and each table is composed of four to six students and one facilitator. The program's objective is not only to teach the content, but to study translanguaging as well; which is the successful incorporation of student's linguistic repertoire into the teaching \cite{translanguage}.

Therefore, given that most students in these environments speak English and Spanish, large datasets must be gathered to train the network since dynamic recognition of both languages is required. Additionally, one's system must be clever enough to address cross-talking and background noise (loud noises, room echoes, etc.) and to tackle the unfavorable microphone-to-speakers ratio as it ranges from 1-to-4 to 1-to-6.

\begin{figure}[h!]
\centering
\begin{subfigure}{.49\textwidth}
  \centering
  \includegraphics[width=.97\linewidth]{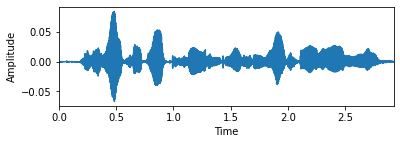}
\end{subfigure}
\begin{subfigure}{.49\textwidth}
  \centering
  \includegraphics[width=.97\linewidth]{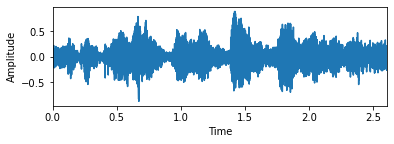}
\end{subfigure}
\caption[Comparison between TIMIT's and AOLME's signals]{Comparison of TIMIT (left) versus AOLME audio (right). On the left, we have a signal of a clear recording from TIMIT. On the right, we have a noisy recording taken from an AOLME conversation. The image on the left pertains to the phrase ``she had your dark suit in greasy wash water all year'', while the image on the right belongs to the phrase ``the computer understands these numbers''.}
\label{fig:wavesIntro}
\end{figure}

To demonstrate the complexity of AOLME audio recordings, refer to Figure \ref{fig:wavesIntro}. The one on the left pertains to the TIMIT dataset, where the phrase ``she had your dark suit in greasy wash water all year'' was read from a book in a quiet environment. On the other hand, the one from the right pertains to the phrase ``the computer understands these numbers''. It was extracted from a conversation that took place in an AOLME classroom.

\begin{figure}[h!]
\centering
\begin{subfigure}{.49\textwidth}
  \centering
  \includegraphics[width=.97\linewidth]{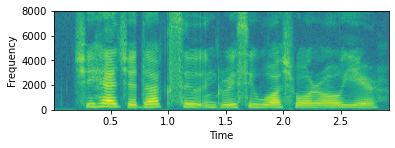}
\end{subfigure}
\begin{subfigure}{.49\textwidth}
  \centering
  \includegraphics[width=.97\linewidth]{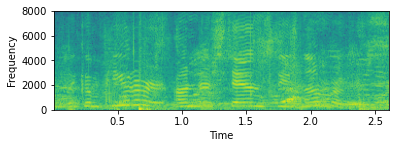}
\end{subfigure}
\caption[Comparison between TIMIT's and AOLME's spectrograms]{Comparison of TIMIT (left) versus AOLME (right) spectrograms. The image on the left pertains to the phrase ``she had your dark suit in greasy wash water all year'' from TIMIT's dataset, and the image on the right belongs to the phrase ``the computer understands these numbers'' from an AOLME conversation.}
\label{fig:specsIntro}
\end{figure}

These two audio plots show that the smoothness between signals varies greatly. The one on the left is less crowded than the one on the right. Additionally, to further portray the differences between noisy and clear recordings, we show the resulting spectrograms of the aforementioned signals in Figure \ref{fig:specsIntro}. Dark blue is associated with silence, while yellow is associated with noise. Notice that the dark blue color is not as abundant and dark on the spectrogram from the right as it is on the spectrogram from the left. Similarly, one can even see some vertical dark blue lines on the spectrogram from the left, representing pauses between sounds. Such pauses are rarely seen in AOLME's recordings.

\section{Thesis Statement}
My thesis is that a phoneme-based method can be effectively used to recognize Spanish and English speech in collaborative learning environments, without the use of language models (LMs) and post-processing techniques. Furthermore, for collaborative learning environments, my thesis is that I can train a low-parameter method that can outperform SOTA methods that utilize larger datasets that do not reflect the complexities of this kind of environment.

\section{Contributions}
This thesis implements a low complexity model capable of training in Spanish and English, individually and jointly. The proposed method is based on DeepSpeech2's \cite{ds2} architecture and trains on Convolutional and Recurrent Neural Networks. Additionally, it uses the CTC loss function to perform labels' predictions of unsegmented sequences.

The thesis uses a simulation framework developed by Antonio Gomez in his dissertation.
The simulation framework uses text transcriptions to produce realistic datasets and allows data augmentation after using multiple random speaker geometries. Based on Pyroomacoustics \cite{pyroom}, this framework lets us expand our datasets by a factor of four.

Furthermore, this thesis generated two different datasets. These were extracted directly from AOLME using ELAN's \cite{elan} and Praat's \cite{praat} software. One is English-based while the other is Spanish-based. The final number of recordings in each dataset is 2,036 and 516 respectively. Recordings ranged from a duration of 0.5 to 5 seconds and each was provided with a respective transcription.

Lastly, the resulting work of this thesis has been submitted and accepted in \cite{ours1}.

\section{Outline}
In the following chapters of this thesis, we discuss SOTA research performed on different areas of Speech Recognition (chapter \ref{sec:background}). In chapter \ref{datasets}, we present the list of datasets used in this work. We include the ones that were available online and the ones produced in this thesis. Later on (chapter \ref{methodology}), we discuss our network architecture and methods used during training. Results are shown in chapter \ref{sec:results} and conclusions, along with future work, are discussed in chapter \ref{sec:conclusions}.

\chapter{Background} \label{sec:background}
At first, this thesis's goal was to recognize English keywords in recordings from collaborative learning environments (AOLME). As research progressed, however, so did the angle. Instead of focusing on one language only, the goal became to recognize two languages (English and Spanish). Additionally, the aim switched from recognizing a set of keywords, to recognizing all phonemes present in the language(s).

In this section, we discuss previous research performed on KWS, E2E speech recognition, and multilingualism. We include descriptions of the datasets used by these systems. We also discuss the research behind E2E speech recognition commercial systems and previous work performed on AOLME by members of the Image and Video Processing and Communications Lab (ivPCL).

\section{Online Datasets} \label{ssec:availableSets}
Datasets that are available online and have been used by different SOTA systems are discussed here. We describe their contents and provide a table (Table \ref{tab:availableSets}) with brief descriptions and access URLs. For this thesis, we used Speech Commands and CSS10 to perform initial assessments of our neural network and used TIMIT for our baseline model.

\begin{table}[h!]
\caption[ASR Datasets Available Online]{Speech recognition datasets that are available online.}
\label{tab:availableSets}
\centering
 \begin{tabular}{| c | c | c |}
 \hline
 \textbf{Name} & \textbf{Description} & \textbf{URL} \\
 \hline\hline
 \multirow{3}{*}{Speech Commands} & Containing 105,829 recordings, it & \multirow{3}{*}{Refer to \cite{sc}} \\
  &  is composed of thousands of repetitions &  \\
  & of 35 different words. & \\
 \hline
 \multirow{3}{*}{TIMIT} & Spoken phrases with phoneme and & \multirow{3}{*}{Refer to \cite{ti}} \\
  &  word level transcriptions. Composed of &  \\
  & 6,300 recordings. & \\
 \hline
 \multirow{4}{*}{CSS10} & Gathered from LibriVox's audiobooks, it & \multirow{4}{*}{Refer to \cite{css10}}\\
  & is composed of 10 different languages. & \\
  & The Spanish subset is composed & \\
  & of 11,111 recordings. & \\
 \hline
 \multirow{4}{*}{EuroNews} & Recordings were extracted from videos & \multirow{4}{*}{Refer to \cite{euronews}}\\
  & that pertained to Euronews portal. & \\
  & Composed of 10 languages and 100 & \\
  & hrs. of data per language. & \\
 \hline
 \multirow{3}{*}{CHiME} & Composed of simulated and real & \multirow{3}{*}{Refer to \cite{chime}}\\
  & data. Real data recorded on  & \\
  & noisy environments. & \\
 \hline
 \multirow{3}{*}{LibriSpeech} & Read speech data based on & \multirow{3}{*}{Refer to \cite{ls}}\\
  & Librivox's audiobooks and composed of & \\
  &  1,000 hrs. of recordings. & \\
 \hline
 \end{tabular}
\end{table}

\subsection{Speech Commands}
Google Speech Commands v2 \cite{sc} is a dataset mainly used for keyword spotting as it is composed of 105,829 single-word utterances. Recordings come from 2,618 speakers and are split into 35 categories (folders); where each category represents a different spoken word. This split is not even, however, since there is a different amount of utterances per word. For the word ``one'' for example, there are 3,890 available utterances, while there are 4,052 for the word ``zero''.

\subsection{TIMIT}
The DARPA TIMIT Acoustic-Phonetic Continuous Speech Corpus is composed of 6,300 read utterances. Spoken by 630 speakers, it is different than most datasets because it offers sentence, word, and phoneme-based transcriptions. A sentence-based example would be the following: "0 39220 Tim takes Sheila to see movies twice a week". Where 0 and 39,220 represent the beginning and end of the recording in terms of the number of samples. Table \ref{tab:timitBreakdown} shows how this sentence is broken down into words and how each word is broken down into phonemes. TIMIT is based on the ARPAbet phoneme set. If stress markers are considered, TIMIT is composed of 52 phonemes, otherwise, it is composed of 39.

\begin{table}[h!]
\caption[TIMIT's Example of Word and Phoneme Transcription]{Few samples of word and phoneme transcriptions for the sentence ``0 39220 Tim takes Sheila to see movies twice a week" of the TIMIT dataset. Values inside the parenthesis represent where each word, or phoneme, starts and ends in terms of the number of samples. For example, the word ``tim'' starts at the sample 2240 and ends at the sample 5540.}
\label{tab:timitBreakdown}
\centering
 \begin{tabular}{| c | c |}
 \hline
 \textbf{Word-Based} & \textbf{Phoneme-Based} \\
 \textbf{Transcriptions} & \textbf{Transcriptions} \\
 \hline
 \hline
 tim (2240, 5540) & t (2240, 2940), ih (2940, 4469), m (4469, 5540)\\
 \hline
 \multirow{2}{*}{takes (5540, 8610)} & tcl (5540, 5860), t (5860, 6570), ey (6570, 8211),\\
 &  kcl (8211, 8610)\\
 \hline
 \multirow{2}{*}{sheila (8610, 14707)} & sh (8610, 11112), iy (11112, 12732), l (12732, 13785)\\
  & ih (13785, 14707)\\
 \hline
 to (14707, 15791) & tcl (14707, 15590), t (15590, 15791)\\
 \hline
 see (15791, 19735) & s (15791, 18882), ix (18882, 19735)\\
 \hline
 ... & ... \\  
 \hline
 \end{tabular}
\end{table}

\subsection{CSS10 (Spanish)} \label{sec:css10}
CSS10 is a corpus composed of speech data from ten different languages. Each of its recordings was obtained from LibriVox audiobooks, along with their respective text transcripts. This research only utilized the Spanish dataset of this corpus, which is composed of 11,111 recordings from three different audiobooks. All recordings were spoken by the same person and their duration ranged from 1 to 23 seconds. It is important to mention that the speaker had a Spanish accent (from Spain) and not an American Spanish accent. More details on section \ref{sec:phonemizer}.

\subsection{CHiME}
CHiME is composed of two subsets, real and simulated. The first one is developed by reading and recording prompts in noisy environments like buses, cafes, pedestrian areas and street junctions. The second one is developed by mixing clear speech recordings with background noise. They are composed of 4,560 and 10,098 recordings, respectively.

\subsection{EuroNews}
The EuroNews Corpus data was gathered from different news of the Euronews portal. Such corpus is split into ten since data was retrieved for ten different languages. The languages are Arabic,  English, French, German, Italian, Polish, Portuguese, Russian, Spanish, and Turkish. For each dataset, around 100 hrs. of videos were taken, with the exception of Polish. Only 60 hrs. were taken for this language.

\subsection{LibriSpeech}
This corpus is a read speech dataset composed of 1,000 hrs. and based on LibriVox's audiobooks. It is split into seven different subsets, dev-clean, test-clean, dev-other, test-other, train-clean-100, train-clean-360, and train-clean-500. The last three pertain to training and are composed of 100, 360, and 500 hrs. of data respectively. The tag ``clean'' represents higher quality and a closer accent to US English. The tags ``dev'' and ``test'' stand for validation and testing datasets, respectively.

\section{Brief Literature Review}
ASR has been a topic of interest for decades. Recent work in areas like KWS, E2E Speech Recognition, and Multilingualism is discussed here. Additionally, we describe the similarities and differences toward this thesis in terms of networks' architectures, datasets, and used resources.

\subsection{Keyword Spotting (KWS)}
The objective in KWS, is to recognize a set of words (usually a short set) in an utterance from those that are not present in the set. Vygon and Mikhaylovskiy \cite{res15} have obtained good results in clean audio. Their approach is triplet loss-based. They use three different inputs, the current training sample, a ``positive'' sample (that shares the same label as the training sample), and a ``negative'' sample (that pertains to a different label). The function's task is to minimize the distance between the training sample and the positive sample and maximize the distance between the training sample and the negative sample. This approach, combined with a k-nearest neighbor classifier (using only 109,000 parameters), achieved a 96.4\% accuracy on the 35-word task of Speech Commands \cite{sc}. 

On the other hand, Berg et al. \cite{kwt3} targeted the same task using multi-head attention and multi-layer perceptron blocks in a transformer encoder. In multi-head attention, a module is run several times in parallel through an attention mechanism. The resulting independent attention calculations are then combined to produce a final score. With this approach, they achieved a 97.69\% accuracy.

The work of Vygon and Mikhaylovskiy is relevant to this thesis since it implements low complexity Convolutional Neural Networks (CNNs) and uses Speech Commands to assess the performance of each model. In contrast, this thesis appends Recurrent Neural Networks (RNNs) on top of the CNNs and uses CTC loss instead of triplet loss. Additionally, this thesis targets phonemes while theirs targets words. In other words, if a word is composed of three phonemes, our model attempts to predict each one of those phonemes while theirs attempts to predict the word.


\subsection{End-to-end (E2E) Speech Recognition}
Unlike KWS, E2E speech recognition is a task that aims to recognize all given characters, phonemes, or words of a given sequence. More specifically, from a set of acoustic features, the objective is to map such features to a label or set of labels. Many attempts have been performed including those of DeepSpeech2 \cite{ds2}, wav2vec 2.0 \cite{wav2vec} and Pytorch-Kaldi \cite{kaldi}. Below, we provide brief descriptions of these methods.

DeepSpeech2 implementation is closest to our approach since its architecture is the backbone of the model proposed in this thesis. It is composed of multiple CNN layers connected to one or more RNN layers. While training, the network uses the CTC loss function to label unsegmented sequences. With this configuration, DeepSpeech2 achieved great results on the LibriSpeech and CHiME datasets. Their results obtained on both datasets are shown in Table \ref{tab:e2eSOTA}.

DeepSpeech2 differs from this work's implementation, however, since they apply a convolution after the RNN layer(s) to filter the total of future context needed. Additionally, this thesis focuses on recognizing IPA phonemes instead of ASCII characters. Furthermore, the augmentation technique used in this work is more complex than theirs. Instead of using a signal-to-noise ratio, we implement an augmentation framework using Pyroomacoustics \cite{pyroom}.

Similar to DeepSpeech2, and our work, wav2vec 2.0 starts with multiple layers of CNNs. In contrast, however, they connect these CNNs to a transformer instead of an RNN. Additionally, instead of using CTC loss, they use contrastive loss, a function that considers a pair of samples and reduces loss by minimizing the distance between similar pairs and maximizing it between dissimilar pairs. Furthermore, wav2vec 2.0 trains with raw waveforms instead of spectrograms and implements LMs for fine-tuning. Their method achieved SOTA results in datasets like LibriSpeech and TIMIT. We provide their results in Table \ref{tab:e2eSOTA}.

Lastly, Pytorch-Kaldi presents a compilation of results on different datasets. They train a vast variety of systems using one or multiple acoustic features. The system that resonates the most with ours is the one they trained with GRUs and tested on the CHiME corpus. The reason being is that they show how GRUs produce better results than vanilla RNNs and LSTMs when it comes to noisy environments. Such results can be seen in Table \ref{tab:e2eSOTA}. Notice that we also show their testing results obtained on other datasets like TIMIT and LibriSpeech.

\begin{table}[h!]
\caption[SOTA E2E results obtained on different datasets]{Speech recognition results from SOTA systems on LibriSpeech, TIMIT, and CHiME. TIMIT's scores are presented in Phoneme Error Rate, while LibriSpeech's and CHiME's in Word Error Rate.}
\label{tab:e2eSOTA}
\centering
 \begin{tabular}{| c | c | c | c | c |}
 \hline
 \multirow{2}{*}{\textbf{System}} & \textbf{TIMIT} & \textbf{LibriSpeech \emph{dev clean}} & \textbf{CHiME \emph{real}} \\
  & \textbf{(PER)} & \textbf{(WER)} & \textbf{(WER)} \\
 \hline\hline
 wav2vec 2.0 & 0.0830 & 0.0210 & - \\
 DeepSpeech2 & - & 0.0515 & 0.2159 \\
 Pytorch-Kaldi & 0.1420 & 0.0620 & 0.1460 \\
 \hline
 \end{tabular}
\end{table}

\subsection{Multilingualism}
Yet another area of ASR that is related to this work is Multilingualism; the recognition of multiple languages even when some might have limited resources (a small number of available recordings for example). In this end, M\"{u}ller et al. \cite{muller1}, \cite{muller2} have shown great results on the Euronews Corpus \cite{euronews} by training networks capable of recognizing two \cite{muller1} and four different languages \cite{muller2}.

Similar to our work, the network architecture of \cite{muller1} and \cite{muller2} is based on DeepSpeech2 as their CNNs connect to RNNs and use the CTC loss function to predict sequences of labels. Additionally, their approach is also phoneme-based as they utilize MaryTTS to translate words into phoneme pronunciations. They also train with mel-based spectrograms, decode with an argmax operation and compute phoneme error rates.

Unlike us, however, they use BottleNeck Features as input to the network and append Language Feature Vectors during training. This so that they can compensate for language-dependent singularities. Additionally, the recordings used to assess their models cannot be considered as noisy as the ones present in AOLME. The reason being is that, even though background music and spontaneous change of language might be present in Euronews, the great quality of the audio and lack of cross-talk makes AOLME a more challenging dataset. Furthermore, while they do not use augmentation, this work uses a simulation and augmentation framework that produces AOLME-like recordings (with cross-talk and background noise).

\section{Commercial Systems} \label{sec:BGcomsyst}
In addition to SOTA open source systems, we explored multiple commercial systems to compare this work's performance on AOLME's recordings against theirs. Using the Spanish version of AOLME-Sentences (section \ref{sec:AOLMEsent}), each recording went through the speech-to-text service of Microsoft Azure, Google Cloud, IBM Watson, and Amazon Web Services. The results obtained from each system, including the one in this research, are shown in section \ref{sec:commSystems}.

In this section, we provide an insight into the network architectures behind each commercial system. We also discuss similarities and differences to this thesis, the number of network parameters, utilized resources, and training data.

\subsection{Microsoft Azure: Speech-to-text} \label{sec:azureSTT}
Although it is not explicitly written, we expect that the network architecture behind Microsoft Azure Speech-to-text is the one presented in Xiong et al. \cite{azure}, where they discuss Microsoft's conversational speech recognition system in detail.

So-called CNN-BLSTM, this new architecture is closely related to the one we propose, since it also connects features extracted from a CNN to one or more layers of bi-directional RNNs. The differences are that this thesis uses GRUs and less than 200 thousand parameters instead of LSTMs and 38 million parameters.

\subsection{Google Cloud: Speech-to-text} \label{sec:googleSTT}
Two of the network architectures mostly used by Google are ``RNN-Transducer'' (RNN-T) \cite{rnnt}, and ``Listen, Attend and Spell'' (LAS) \cite{las}. Even though it is not written, we expect one of these two is the backbone of Google Cloud's Speech-to-text service.

The first one, RNN-T, is mainly used for streaming ASR \cite{streamingrnnt} \cite{mobilernnt} and is composed of three networks: an encoder, a prediction, and a joint network. Both, the encoder and prediction networks, are trained independently and are composed of multiple LSTM layers. Their outputs are then combined through the joint network (usually involving a forward-backward algorithm with a softmax function) that in return predicts a distribution of the next label.

On the other hand, the LAS architecture is composed of multiple pyramidal bi-directional LSTMs that connect to an attention-based LSTM transducer. SpecAugment \cite{specaugment} uses LAS architectures to experiment with techniques like time warping, frequency masking, and time masking, for data augmentation. In \cite{las2}, the authors optimize the LAS architecture by incorporating multi-head attention and hypothesize that this adjustment results in better performance when dealing with noisy data.

\subsection{IBM Watson: Speech-to-text} \label{ibmSTT}
Multiple documents in \cite{ibmSTT} state that IBM's Speech-to-text service is based on bidirectional LSTMs using the CTC loss function. Audhkhasi et al. \cite{ibmlstm1} explored the technique of training an acoustic to word (A2W) model, thereby challenging the previous acoustic to phoneme, or character, models. This technique by itself, however, did not perform well. To alleviate this, they trained their A2W model on top of a phoneme-BLSTM system and used global vectors word embeddings to initialize the last fully connected layer. Their A2W model consisted of 5 BLSTM layers, a fully connected layer and the softmax activation function.

Later on, Audhkhasi et al. \cite{ibmlstm2} re-evaluated the aforementioned system and obtained better performance on the A2W model. They sorted their data in ascending order, added delta and delta-delta coefficients, incorporated Nesterov's momentum-based stochastic gradient descent, used projection layers on the outputs, and increased the size of their phone BLSTM system. However, even though better results were obtained, the out-of-vocabulary (OOV) issues remained; the system was not able to recognize words that were not in the vocabulary. To solve this, they trained a BLSTM network using characters and words jointly. Such a tactic ameliorated the performance of OOV words.

\subsection{Amazon Web Services: Amazon Transcribe}
Similar to the previous commercial systems, to our best knowledge, the network architecture behind Amazon Transcribe (Amazon's Speech-to-text service) is that of an RNN-T. A system composed of three different networks (as discussed in \ref{sec:googleSTT}), where the first two are trained independently and come together through the third one to predict the next output label. Amazon has recently researched this architecture in \cite{minimumrnnt} and \cite{subwordrnnt}.

On the first one, the minimum word error rate (MWER) method is appended during training to the RNN-T. By doing so, the network ``smartly groups alignments of the same hypothesis together'' \cite{minimumrnnt}. On the second one, subword regularization is studied using an RNN-T to analyze and improve the performance of words unseen by the network. \cite{subwordrnnt}.

\section{Prior Work on AOLME}
Different members of ivPCL have performed extensive research in the Collaborative Learning Environments of AOLME. Recently, Teeparthi et al. \cite{sravani} \cite{ours4} developed a fast and effective system that detects objects, and tracks hands and keyboards in long videos of collaborative learning environments. This was achieved by incorporating different object detection methods like projections, clustering, and tracking.

Also in the area of detection, Darsey \cite{sravani8} studied hand movement detection using histograms of optical flow, patch-color classification, and space-time patches of video. Shi et al. \cite{sravani38}, \cite{sravani39} and Tapia et al. \cite{sravani41} focused their research on head and face detection, respectively. The first one not only focused on head detection, but also the detection of the subject's attention. By looking at the subjects' face direction and using AM-FM models, both tasks were achieved. The latter used low complexity CNNs to study face detection and determine the advantages of using AM-FM representations instead of raw images.

Furthermore, the studies of Jacoby et al. \cite{sravani22}, \cite{sravani23} and Eilar et al. \cite{sravani11}, \cite{sravani12} examined the topic of human activity. Jacoby et al. focused on the study of writing, typing, and talking, while Eilar et al. proposed an open-source system capable of detecting human activity present in video recordings. Similarly, Shi et al. studied human activity in these kind of environments but focused only on talking detection \cite{ours2} and person detection \cite{ours5}.

Lastly, in terms of detection, Jatla et al. \cite{sravani24}, \cite{sravani25}, implemented a SOTA automated segmentation technique in image processing models that automate the detection of coronal holes while Tran et al. \cite{ours3} studied facial recognition.

Other fields aside from detection have also been studied. For example, Ulloa et al. \cite{sravani6} proposed a 3D CNN system that assists cardiologists, as it is capable of predicting one-year all-cause mortality by analyzing echocardiographic videos. Carranza et al. \cite{sravani5} were able to maximize throughput with optimized 2D-FFT libraries and vector-based memory I/O. Kent et al. \cite{sravani28} incremented speed on 2x2 and 3x3 max-pooling with the help of FPGAs.

\chapter{Datasets}\label{datasets}
Datasets that were developed in this research are discussed here, as well as details on their usage, contributions, and development. We also provide additional information on Speech Commands and the way their data is split into training, validation and testing sets.

\section{Speech Commands Details} \label{datasets:sc}
As mentioned in \cite{sc}, this dataset is split into training, validation, and testing using a hashing function. Such a function allows them to append additional files to future versions of the dataset and maintain independence between sets. 

In \cite{sc}, they provide two documents, \emph{validation\_list.txt} and \emph{testing\_list.txt}. The number of recordings listed in each file is 9,981 and 11,005, respectively. Therefore, about 10\% of the recordings are kept for validation, another 10\% for testing, and the remaining 80\% for training.

Lastly, although it is not mentioned, the distribution of words between sets seems proportional. For example, for the word ``right'', the validation set contains 363 samples, the testing set 396 and the training set the remaining 3,019 samples.

\section{Revision of CSS10 (Spanish)} \label{datasets:css10}
Before utilizing the CSS10 dataset, multiple adjustments were performed to the text transcripts. For instance, any digits or uppercase letters present were translated to written numbers and lowercase letters, respectively. Roman numerals were translated as well; i.e., ``Carlos VII" was translated to ``Carlos séptimo", where séptimo stands for seventh in English. Any special character (`?', `:', `.', `,', `-', `¡', among others) was removed from the transcripts. In the end, the resulting transcripts were composed only of characters in the Spanish alphabet (including letters with accents) and the white space character.

Additionally, for this work, only a subset of the full dataset was used. If a recording included one of the words ``Cero'', ``Uno'', ``Dos'', ``Tres'', ``Cinco'', ``Número'' or ``Números'', the recording was kept; otherwise, it was discarded. The reason being is that initially, this work had a keyword spotting angle. Nonetheless, the resulting dataset of 577 recordings was not modified. Each recording was simulated and augmented, as described in section \ref{sec:pyroom}, to mimic real recordings of AOLME's classrooms.

After simulation and augmentation, the resulting dataset was composed of 2,308 recordings. Such were split into training and testing to assess the performance of the low-complexity model proposed in this document. Results of this assessment can be found in section \ref{sec:resultsCSS10}.

\section{AOLME Datasets} \label{sec:aolmeSets}
AOLME's after-school program has been hosted in six different terms. Each term is composed of approximately ten different groups. On average, groups have ten sessions each term and each session lasts around one and one and half hours. With these numbers, one can say that AOLME is composed of approximately 900 hours of recordings from Collaborative Learning Environments.

In this research, three different datasets were developed using the aforementioned corpus. As described below, the first one is composed not only of recordings extracted from AOLME but also of recordings from Speech Commands and TIMIT. The second is composed of spoken sentences uniquely extracted from AOLME. The third one was developed using AOLME's transcripts rather than its recordings.

\subsection{AOLME-Numbers} \label{sec:aolmeNums}
Three different datasets were used to develop this subset. Speech Commands was used for training while TIMIT and AOLME were used for testing. Three different categories were considered: ``One'', ``Zero'' and ``Others''. Each category was composed of single-word recordings. If the spoken word was not ``zero'' or ``one'', it was added to the ``others'' category.

The training set was composed of 1,500 recordings. Five hundred were obtained from the ``zero'' folder of Speech Commands. Other 500 were obtained from the ``one'' folder and the remaining 500, for ``others'', were randomly chosen from the remaining 33 folders. The two testing datasets were obtained by extracting spoken words from AOLME and TIMIT. A Python script was utilized to perform such a task. The script would play a selected recording and ask the user for the start and end times of the word of interest (a sub-sample of the recording). It would then use these times to crop and play the sub-sample. Then, the user would have the option of saving or denying the sub-sample (if denied, the script would ask for the times once again). This would happen recursively until a satisfying sub-sample was obtained. Table \ref{tab:aolmeNums} shows the number of utterances (sub-samples) used from each dataset for each category.

\begin{table}[h!]
\caption[AOLME-Numbers Dataset]{Number of utterances used from each dataset for the categories ``One'', ``Zero'' and ``Others''.}
\label{tab:aolmeNums}
\centering
 \begin{tabular}{|c | c  c  c|}
 \hline
 \textbf{Dataset} & \textbf{``One''} & \textbf{``Zero''} & \textbf{``Others''} \\
 \hline\hline
 Speech Commands & 500 & 500 & 500 \\ 
 TIMIT & 10 & 10 & 10 \\
 AOLME & 12 & 12 & 12\\
 \hline
 \end{tabular}
\end{table}

This subset was used to perform a preliminary assessment of DeepSpeech's performance on AOLME's noisy recordings.

\subsection{AOLME-Sentences} \label{sec:AOLMEsent}
AOLME-Sentences was developed by extracting spoken sentences from different AOLME recordings. Two versions were considered in this work, one for Spanish and one for English. Sessions from six different groups were used to maintain diversity in gender and the number of speakers; three groups for Spanish and four for English, where one overlapped between the two of them.

\begin{figure}[h!]
 \centering
 \includegraphics[width=0.8\columnwidth]{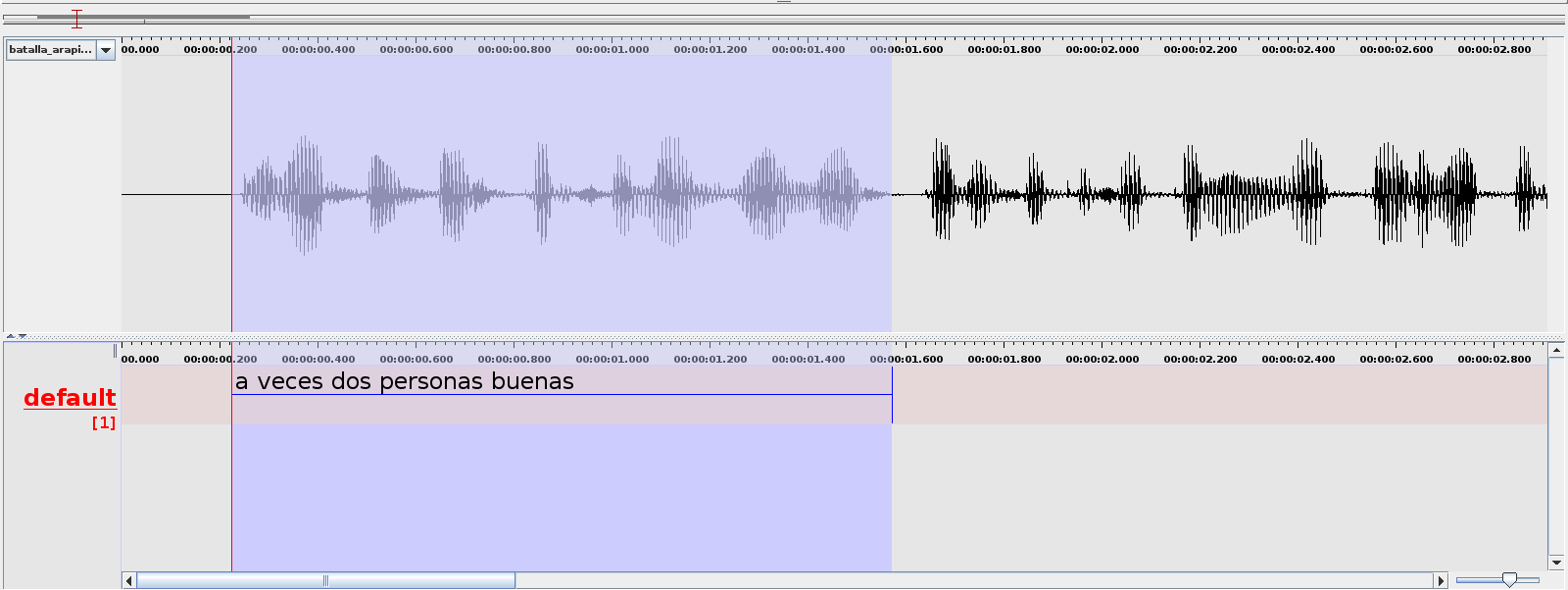}
    \caption[ELAN's Graphical User Interface]{Example of Elan's GUI. Notice the signal on top and the text transcription below.}
    \label{fig:elanUI}
\end{figure}

Initially, the same script described in \ref{sec:aolmeNums} was used to extract these utterances. However, its implementation was tedious and lengthy because it required its entries to have a three-digit precision. Thankfully, a tool was available to help alleviate this process, ELAN Software \cite{elan}. Its GUI is shown in Figure \ref{fig:elanUI}. One can see that after uploading a recording to the program, the signal of the recording is placed above an annotation timeline. Thus, instead of inputting times of up to three decimal points, one can simply choose a time-frame with the mouse and type the transcription right after. The transcription will then be saved underneath in the annotation timeline (as shown in the figure).

\begin{figure}[h!]
 \centering
 \includegraphics[width=0.8\columnwidth]{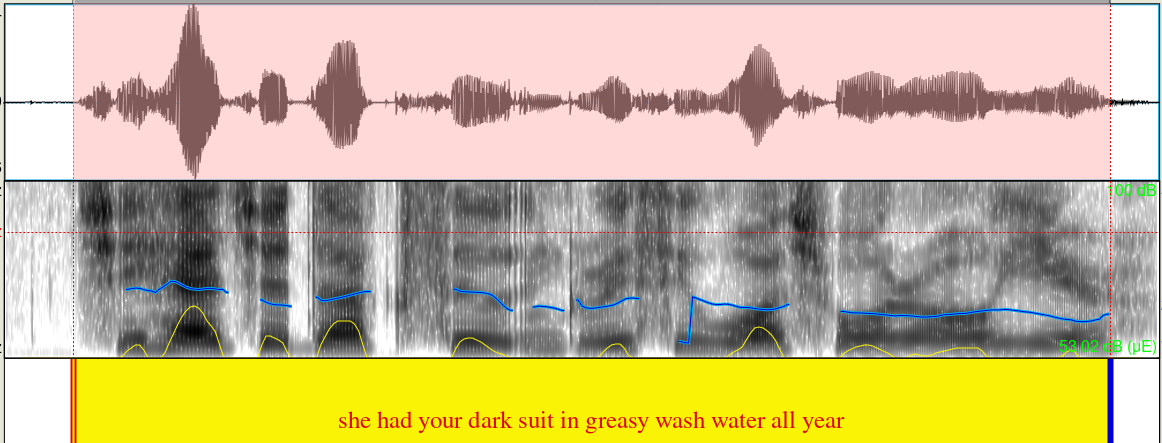}
    \caption[TIMIT's Recording in Praat's GUI]{Example of Praat's GUI using a recording from TIMIT. The phrase ``she had your dark suit in greasy wash water all year" pertains to this recording.}
    \label{fig:praatTimit}
\end{figure}

\begin{figure}[h!]
 \centering
 \includegraphics[width=0.8\columnwidth]{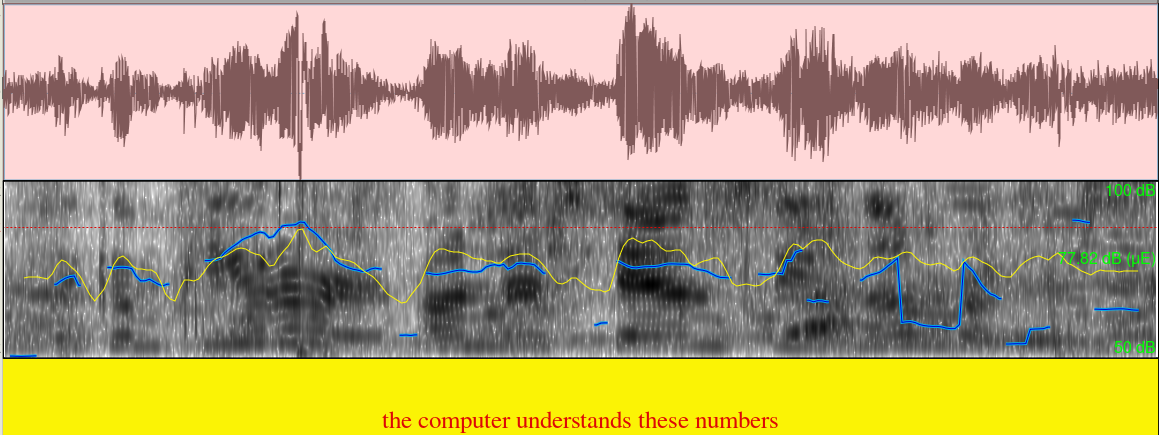}
    \caption[AOLME's Recording in Praat's GUI]{Example of Praat's GUI using a recording from AOLME. The phrase ``the computer understands these numbers" pertains to this recording.}
    \label{fig:praatAolme}
\end{figure}

Although this tool was helpful, it was not sufficient. The amount of noise in AOLME's recordings complicated the task of picking accurate time-frames from only looking at the wave signal. A new program called Praat \cite{praat} was then found as the solution. This software not only displays the recording's signal but its spectrogram as well. Additionally, it also allows the user to display the Pitch, Intensity, Pulse, or Formants of the recording. Two examples of its GUI can be seen in Figures \ref{fig:praatTimit} and \ref{fig:praatAolme}. Figure \ref{fig:praatTimit} is a recording from TIMIT of the phrase ``she had your dark suit in greasy wash water all year''. Figure \ref{fig:praatAolme} is a recording from AOLME of the phrase ``the computer understands these numbers''. 

In these two figures, from top to bottom one can see the recording's signal, the spectrogram, and the text transcription. The blue and yellow lines inside the spectrogram represent the Pitch and Intensity, respectively. If one pays close attention, one can see that TIMIT's Pitch and Intensity are somewhat organized, while AOLME's are all over the place. Therefore, as mentioned in section \ref{intro:aolme}, if both figures are compared against each other, one can see the difference between clear and noisy audio.

Praat simplifies the task of annotating recordings because one can focus on multiple features when picking time-frames. If the wave signal is noisy and does not give a clear idea of where a person has stopped talking, one can use the spectrogram, intensity, pitch, pulse, or formants to make a more precise guess.

In the end, a total of 516 utterances were extracted for Spanish and 2,036 for English. Recording's duration varied from 0.5 seconds to 5 seconds. Table \ref{tab:aolmeutts} shows the number of recordings extracted from each group for Spanish and English. It also provides the utterances' duration of each group in minutes. Notice that a letter has been given to each group, so that anonymity is kept; and that group `B' is the one that overlaps between both languages.

\begin{table}[h!]
\caption[AOLME-Sentences]{Number of utterances extracted from each group for Spanish and English.}
\label{tab:aolmeutts}
\centering
 \begin{tabular}{|c c c | c c c|} 
 \hline
 \multicolumn{3}{|c|}{\multirow{2}{*}{\textbf{Spanish}}} & \multicolumn{3}{c|}{\multirow{2}{*}{\textbf{English}}}\\
 \multicolumn{3}{|c|}{} & \multicolumn{3}{c|}{}\\
 \hline
 \multirow{2}{*}{\textbf{Group}} & \multirow{2}{*}{\textbf{Utterances}} & \textbf{Duration} & \multirow{2}{*}{\textbf{Group}} & \multirow{2}{*}{\textbf{Utterances}} & \textbf{Duration} \\
  &  & \textbf{(minutes)} & & & \textbf{(minutes)} \\
 \hline\hline
 A & 219 & 2.23 & B & 677 & 6.48 \\ 
 B & 158 & 2.24 & D & 181 & 3.06 \\
 C & 139 & 1.79 & E & 1,087 & 6.40\\
 - & - & - & F & 91 & 1.50 \\
 \hline
 \end{tabular}
\end{table}

The Spanish version of this dataset was used as a test set to assess the performance of the low complexity model proposed in this work. It was also used as a test set to assess the performance of the commercial systems discussed in section \ref{sec:BGcomsyst}. Results from both assessments are provided in section \ref{sec:commSystems}. The English version of this dataset is yet to be used, as it is mentioned in chapter \ref{sec:conclusions}.

\subsection{AOLME-Transcripts} \label{sec:aolmeTranscr}
In addition to recordings of each session, AOLME also possesses transcripts for some of these recordings. Examples of these transcripts are shown in the first column of Table \ref{tab:origTranscripts}. This column shows that in addition to what each speaker said, the transcripts include speakers' pseudonyms, speakers' actions, translations to English (if the speaker spoke in Spanish), and comments on current situations. Therefore, a first correction was applied to the original transcripts, since there was no interest in keeping this additional information. 

Another Python script was used to perform such a task. The results of this script are shown in column two of the table. Notice that in row one, the speaker's pseudonym and action were removed from the original transcript. Similarly, in row two, the pseudonyms of the speaker and the ``inaudible'' comment was removed. Lastly, in row three, the pseudonyms of the speaker and the translation to English was removed.

Once this additional information was removed, a second correction took place. Similar to CSS10 (section \ref{datasets:css10}), original transcripts contained uppercase letters, digit-based numbers and special characters. Therefore, the same rules were applied to remove special characters, and replace uppercase letters and digit-based numbers. Column three shows the resulting transcripts after applying the aforementioned rules. In the given examples, notice that uppercase letters were replaced, and that question marks and dots were fully removed from the sentences.

\begin{table}[h!]
\caption[Examples of AOLME-Transcripts]{Relationship between original transcripts of AOLME and the adjustments performed in order to produce AOLME-Transcripts.}
\label{tab:origTranscripts}
\centering
 \begin{tabular}{|c | c | c |}
 \hline
 \multirow{2}{*}{\textbf{Original Transcript}} & \multirow{2}{*}{\textbf{First Correction}} & \multirow{2}{*}{\textbf{Second Correction}} \\
  & & \\
 \hline\hline
 Miguel: Hello? [Miguel answers & \multirow{2}{*}{Hello?} & \multirow{2}{*}{hello}\\
 his mom on the phone] & & \\
 \hline
 Facilitator: (Inaudible). & \multirow{2}{*}{You know?} & \multirow{2}{*}{you know}\\
 You know? & & \\
 \hline
 Matias: Ahí esta mejor. & \multirow{2}{*}{Ahí está mejor.} & \multirow{2}{*}{ahí está mejor}\\
 (This is better.) & & \\
 \hline
 \end{tabular}
\end{table}

After these rules were applied, the resulting transcripts were split into Spanish and English. Reason being is explained in-depth in section \ref{sec:phonemizer}. Additionally, if a sentence included both languages, it was removed from the dataset. At the end, the Spanish and English versions of this dataset were composed of 2,857 and 1,096 sentences, respectively.

Afterwards, these sentences were run through a speech synthesizer, thereby generating recordings with AOLME's classroom context. Furthermore, these recordings were simulated, so that they would mimic AOLME's real recordings, including cross-talk and background noise. Lastly, data augmentation was performed in order to increase the size of the AOLME-Transcripts dataset. Details on how this dataset was synthesized, simulated and augmented are provided in sections \ref{sec:ttsmp3} and \ref{sec:pyroom}.

Although both versions, Spanish and English, were synthesized, only the Spanish version was simulated and augmented. As listed in chapter \ref{sec:conclusions}, the English version will be simulated and augmented in the future. The resulting number of recordings for the Spanish AOLME-Transcripts dataset was 37,264. Such recordings were used to train and test the low complexity network proposed in this work. Validation and testing results are provided in section \ref{results:AOtranscr}.

It is important to mention that the transcripts used to populate AOLME Transcripts were chosen carefully. They do not overlap with transcripts that pertained to the sessions' recordings of AOLME-Sentences. This way, independence between datasets is guaranteed.

\section{Data Generation}
This section explains the procedures taken to synthesize, simulate or augment certain datasets used in this thesis.

\subsection{Speech Synthesis} \label{sec:ttsmp3}
To generate audio from text, a text-to-speech service was utilized. Powered by Amazon Polly, TTSMP3 \cite{ttsmp3} accepts a piece of text (paragraph, phrase or word) as input and outputs a recording of such text. It offers different languages, as well as different voices to pick from each language. Once the text is inputted and the language-voice pair is chosen, it generates a recording in such language with the chosen voice. For Spanish, all four available voices were used, Lupe, Penelope, Miguel and Mia. For English, only three voices have been utilized so far, Ivy, Joey and Justin. The rest of the English voices will be used in the future.

To synthesize an audio, one has to manually type the text in the website, select the language, choose the voice, and click download. Performing such task manually would have been tedious and time-consuming. Furthermore, if one wanted access to the API in hope of automatizing this task, one would have had to spend \$99.00 for a one-year subscription. Therefore, to save time and money, a different method was used.

Such a method was performed with the aid of Selenium WebDriver \cite{selenium}, a framework that allows users to automatically perform quality assurance assessments in their web applications. For instance, it can automate a website's commands like clicking, selecting, typing, and inserting text. Such commands and functionalities were used in this research to automate the speech synthesis task mentioned above.

\begin{figure}[h!]
 \centering
 \includegraphics[width=0.8\columnwidth]{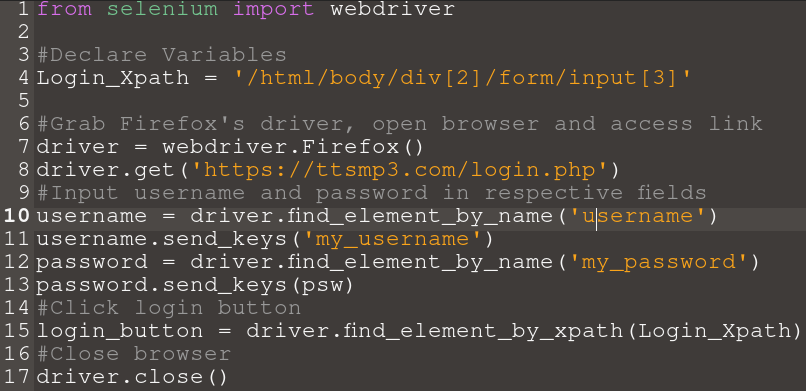}
    \caption[Python example of Selenium's API]{Python example of Selenium's API.}
    \label{fig:seleniumCode}
\end{figure}

Sample commands of Selenium's Python API are shown in Figure \ref{fig:seleniumCode}. Lines 7 and 8 use Firefox's driver to open the browser and access the link given to the \emph{get} method. Line 10 locates the username field in the web-page using the ``name'' attribute. Line 11 uses the method \emph{send\_keys} to type the username in this field. In other words, lines 10 and 11 mimic a user clicking inside the username field and inputting the username. Lines 12 and 13 repeat the aforementioned process, but for the password. Line 15 clicks the login button by using its ``xpath'' atrribute, while line 17 closes the browser. In summary, the script shows how one can open a browser, access a link, type the username and password, click the login button, and close the browser. A complete rundown of the commands used to automate this task can be found in Appendix \ref{appendices:selenium}.

Given that the Spanish version of AOLME-Transcripts was composed of 2,857 sentences, after running them through TTSMP3, using each Spanish voice mentioned above, a total of 11,428 recordings were produced. These were checked thoroughly, since some recordings did not sound human-like. They either sounded ``robotic'' or a mispronunciation of a word was present. After this quality assurance step, 9,316 recordings were kept.

Although all 1,096 English sentences were also run through TTSMP3 (using three different voices), a thorough check has not yet been performed. This will be performed in the future as stated in chapter \ref{sec:conclusions}.

\subsection{Data Augmentation and Simulation} \label{sec:pyroom}
The next step in creating a dataset that would resemble original recordings of AOLME was to use the resulting recordings from TTSMP3 and run them through a simulator. This work used Pyroomacoustics \cite{pyroom} (Pyroom) to achieve this.

\begin{figure}[h!]
\centering
\begin{subfigure}{.49\textwidth}
  \centering
  \includegraphics[width=.97\linewidth]{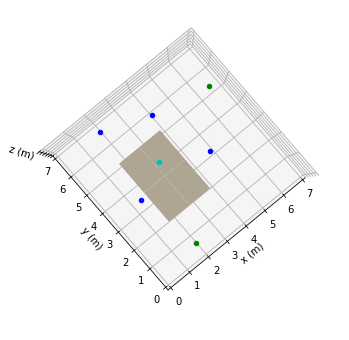}
\end{subfigure}
\begin{subfigure}{.49\textwidth}
  \centering
  \includegraphics[width=.97\linewidth]{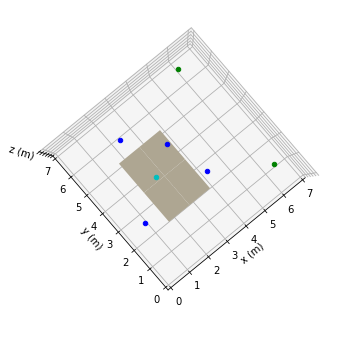}
\end{subfigure}
\begin{subfigure}{.49\textwidth}
  \centering
  \includegraphics[width=.97\linewidth]{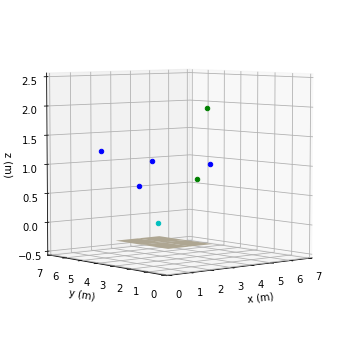}
\end{subfigure}
\begin{subfigure}{.49\textwidth}
  \centering
  \includegraphics[width=.97\linewidth]{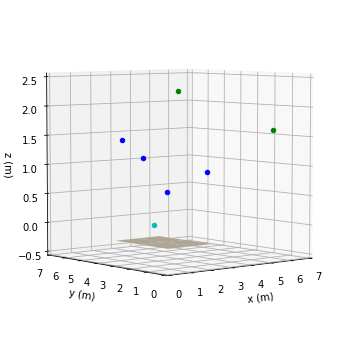}
\end{subfigure}
\caption[Pyroomacoustics: Examples of Geometries]{Examples of different geometry configurations on Pyroomacoustics. The two images on the left represent the same configuration while the images on the right represent a different one. The images on the top represent the top-view of the classroom and the images on the bottom represent a side-view.}
\label{fig:pyroomDrawings}
\end{figure}

Pyroom allows the setup of a room, with multiple speakers talking and one or more microphones recording. Using approximate dimensions of AOLME's classrooms and student's locations, multiple geometries were attempted. Two examples are shown in Figure \ref{fig:pyroomDrawings}. The two images on the left pertain to the same configuration; a top-view and a side-view. Similarly, on the right, we show a different configuration where each dot has been moved to a different position. Notice that both examples resemble a table of an AOLME classroom. Each source (blue dot) represents a different student, each ``room noise'' (green dot) represents a different background noise and the turquoise dot represents the microphone.

Both examples are different from each other in the sense of geometries. Notice that the position of each speaker changes from the images on the left to the images on the right. Similarly, the microphone and room noises are placed in different locations.

On Pyroom, each source (blue dots) reproduces a different utterance, so that cross-talking is mimicked. Similarly, each room noise (green dots) is provided with noisy recordings to resemble loud noises in the classroom. The objective is to carefully configure pitch and gain attributes of each source and each noise so that a certain source is captured by the microphone. In other words, once a source is ``chosen'' by the user, the microphone's task is to recognize this source's recording. Therefore, the recordings that were obtained from TTSMP3 are the ones that we provided to this source. All other sources were provided with random recordings.

\begin{figure}[h!]
\centering
\begin{subfigure}{.32\textwidth}
  \centering
  \includegraphics[width=.97\linewidth]{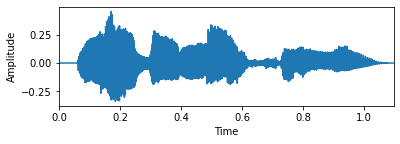}
\end{subfigure}
\begin{subfigure}{.32\textwidth}
  \centering
  \includegraphics[width=.97\linewidth]{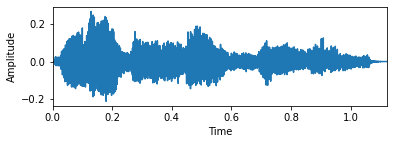}
\end{subfigure}
\begin{subfigure}{.32\textwidth}
  \centering
  \includegraphics[width=.97\linewidth]{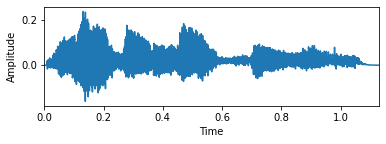}
\end{subfigure}
\caption[Pyroomacoustics: Original vs. Simulated Signals]{Difference between original and simulated signals. The original signal is on the left, while simulated signals are in the middle and right.}
\label{fig:pyroomWaves}
\end{figure}

Figure \ref{fig:pyroomWaves} shows the outcome of simulating an utterance using Pyroom. In the figure on the left, one can see the TTSMP3 synthesized recording of the Spanish phrase ``nada más uno'' (which translates to ``only one''). On the middle and the right, one can see resulting recordings from Pyroom using two different geometry configurations. Notice that not only are they different from the original signal, but they are also different from each other.

Therefore, not only did Pyroom permit simulations to closely resemble recordings from AOLME, but it also allowed each TTSMP3 recording to be augmented. However, to generate realistic simulations that were not too similar, only four different Pyroom geometries were configured. As a result, the Spanish dataset of AOLME Transcripts was expanded by four and resulted in a total of 37,264 recordings.

\chapter{Methodology} \label{methodology}
In this section, four main concepts are discussed. First, we describe the proposed network architecture of this thesis. Second, we discuss the method used to convert ASCII-based text into sequences of IPA phonemes. Third, we explain the learning rate scheduler used in this thesis. Lastly, we cover a baseline model which compares a SOTA open-source speech recognizer to a custom SVM-PCA classifier.

\section{Network Architecture}\label{sec:netArch}
\begin{figure}[h!]
 \centering
 \includegraphics[width=0.7\columnwidth]{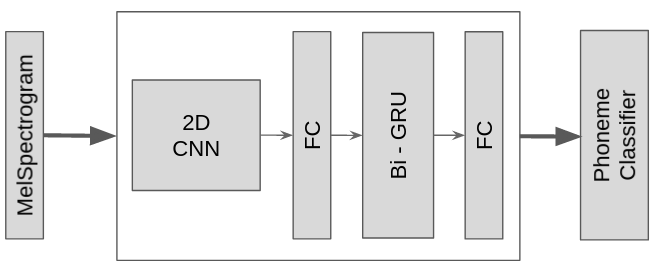}
    \caption[Low-Complexity Network Architecture]{Diagram of the network architecture for the low-complexity model proposed in this work.}
    \label{fig:netArch}
\end{figure}

The low complexity model proposed in this research was built on PyTorch. A complete overview of our network architecture is provided on Python in Appendix \ref{appendices:network}. Additionally, a sample code demonstrating how to run the network is provided in \ref{appendices:train}. A complete implementation of our code is provided in \cite{ournetwork}. Our model is motivated by the architecture of DeepSpeech2, and it was built on top of Michael Nguyen's \cite{assemblyai} implementation. We provide a diagram of the proposed architecture in Figure \ref{fig:netArch}.

First, mel-spectrograms of each recording were calculated and used as input. Before calculating the spectrograms, however, each utterance was zero-padded if it was shorter than 0.3 seconds. Spectrograms were calculated using the default values of PyTorch's \emph{MelSpectrogram} transformation (40ms window size, 20ms hop length, and 128 mel filters).

Most datasets' recordings were mono (one-channel) in the \emph{WAV} format using a \emph{LINEAR16} encoding and a 16,000 Hz sample rate. If recordings were not configured in this format, they were carefully converted using \emph{ffmpeg}. The command used for such a task is provided in Appendix \ref{appendices:ffmpeg}.

After spectrograms were calculated, they were passed through a 1-layer 2D CNN and a fully connected layer right after. For the CNN, stride value and kernel size never changed. They remained 2-by-2 and 3-by-3, respectively, in all runs. Five different values were considered for the number of filters in the CNN, 4, 8, 12, 16, and 32. The number of input units for the fully connected layer was calculated dynamically as shown in the code for the variable \emph{in\_feats}.

Layer normalization was applied next and the GELU activation function right after. A set of bi-directional GRUs followed the activation function. Four different numbers were considered in terms of number of GRU layers, 2, 3, 4 and 8. For input and hidden units, four different pairs were considered, 32-32, 32-64, 64-64 and 64-32.

At the end, two fully connected layers were implemented with a GELU activation function and a dropout in between. The outputs of the network were then run through a logarithmic softmax function.

The Connectionist Temporal Classification (CTC) loss function was used to transform these outputs into a conditional probability distribution of labels. To obtain a resulting sequence of phonemes, an \emph{argmax} decoder was utilized. Additionally, from the resulting sequences, blank labels and repeated labels were removed from such sequences. To illustrate this step, we provide Table \ref{tab:ctcoutputs}. On the left, we show the words that are being predicted (ground truth). In the middle, we provide possible outputs of the aforementioned decoding method. Notice that some phonemes might be repeated and some $blank$ labels (represented as underscores) are present in the prediction. Therefore, after removing them, predictions like the ones in the third column are obtained.

\begin{table}[h!]
\caption[Network's outputs using CTC and argmax]{Examples of predictions produced by the network using CTC and argmax. Original words are provided in the first column. Predicted sequences of phonemes are provided in the second column, and corrected predictions are provided in the third column. Notice that the $blank$ label is represented by an underscore.}
\label{tab:ctcoutputs}
\centering
 \begin{tabular}{|c | c | c|} 
 \hline
 \multirow{2}{*}{\textbf{Word to be Predicted}} & \multirow{2}{*}{\textbf{Prediction in Phonemes}} & \textbf{Prediction After} \\
 & & \textbf{Corrections} \\
 \hline\hline
 audio & ɔː d \_ ɪ oʊ oʊ oʊ & ɔː d ɪ oʊ \\
 network & n ɛ \_ \_ t w \_ ɜː k k & n ɛ t w ɜː k \\
 architecture & ɑːɹ ɑːɹ k ɪ \_ t ɛ k tʃ \_ ɚ & ɑːɹ k ɪ t ɛ k tʃ ɚ \\
 simulation & s ɪ m j ʊ l eɪ eɪ eɪ ʃ ə ə n \_ & s ɪ m j ʊ l eɪ ʃ ə n \\
 work & \_ \_ w ɜː ɜː k & w ɜː k \\
 \hline
 \end{tabular}
\end{table}

Once the blank and repeated labels were removed, the \emph{Levenshtein distance} operation was performed to obtain Phoneme and Word error rates. This operation determines how similar one string is to another by keeping track of the number of insertions, deletions, and substitutions. An insertion occurs when a label is added to a prediction. For example, if the word to predict is ``shine'' and the prediction is ``shrine'', the `r' character is considered an insertion. On the other hand, a deletion occurs when a character is omitted in the prediction. The prediction ``sin'' from the word ``sing'' would be a deletion since the letter `g' is omitted. Lastly, an addition happens when a label is added to the prediction. If the word ``tree'' is the word to predict and ``three'' is the prediction, the character `h' is the addition. For further details, refer to \cite{Levenshtein}.

It is important to mention that the GELU activation function was used throughout all runs. Other activation functions are yet to be considered. Additionally, all runs were performed on System76's operating system Pop!\_OS 20.04 LTS and were run on a single 8GB GeForce GTX 1050 Mobile Nvidia GPU.

\section{Phonemization}\label{sec:phonemizer}
Given that the goal is to recognize two different languages simultaneously (Spanish and English), this work proposes the usage of an extended character set based on the International Phonetic Alphabet (IPA). By doing so, phonemes from both languages that sound the same can be identified with the same phoneme (label).

Table \ref{tab:phonemizerWords} shows examples of \emph{phonemized} words in both languages. Notice that the phonemes that share the same sound in both languages (for particular words) have been bolded. For example, row one shows that the phoneme `b' sounds the same for the Spanish word ``babear'' and the first `b' of the English word ``barbecue''. Similarly, row two shows how phonemes `m' and `s' sound the same in the words ``museo'' and ``mouse'', for Spanish and English respectively. 

\begin{table}[h!]
\caption[Similar Phonemes Between Spanish and English]{Examples of words demonstrating phoneme similarity between Spanish and English. Notice that column one provides an English translation for each Spanish word.}
\label{tab:phonemizerWords}
\centering
 \begin{tabular}{|c c | c c|} 
 \hline
 \multicolumn{2}{|c|}{\textbf{Spanish}} & \multicolumn{2}{c|}{\textbf{English}}\\
 \hline
 \textbf{Word in ASCII} & \textbf{Word in IPA} & \textbf{Word in ASCII} & \textbf{Word in IPA}\\
 \hline\hline
 babear (drool) & \textbf{b} a β e a ɾ & barbecue & \textbf{b} ɑːɹ b ɪ k j uː \\
 museo (museum) & \textbf{m} u \textbf{s} e o & mouse & \textbf{m} aʊ \textbf{s} \\ 
 hola (hello) & o \textbf{l} a & hello & h ə \textbf{l} oʊ \\ 
 carro (car) & \textbf{k} a ɾ ɾ o & cat & \textbf{k} æ t \\ 
 baile (dance) & b \textbf{aɪ} l e & time & t \textbf{aɪ} m \\
 \hline
 \end{tabular}
\end{table}

These translations to IPA were performed using bootphon's Phonemizer library \cite{phonemizer}. Appendix \ref{appendices:phonemizer} shows a sample Python script of its usage. Briefly, the user provides the code with a language and backend to use. Then, the user determines which word or phrase (in the language of interest) will be phonemized. Then, one runs the phonemizer and receives a set of IPA phonemes equivalent to the given text. This thesis used the backend ``espeak'' since it supports IPA phonemes. More information regarding this backend and its phonemes' transcriptions can be found here \cite{espeak}.

As initially described in section \ref{sec:aolmeTranscr}, AOLME-Transcripts had to be split in two, Spanish and English. The reason being is that if a word in one language is phonemized using a different language (in the phonemize), the produced set of phonemes is erroneous. For example, take the Spanish word ``carro''. Phonemizing this word with the English language would output ``k ɑːɹ ɹ oʊ'', which is nothing similar to its translation in Table \ref{tab:phonemizerWords} (which was performed with the Spanish language). Instead, it seems as if the Spanish word ``carro'' rhymes with the English word ``sparrow'', which translates to ``s p æ ɹ oʊ'', which clearly, should not be the case.

In other words, translating a Spanish word to phonemes, using the English language would do exactly that; it will phonemize the word with an English accent. The same would happen if one tries to phonemize an English word using the Spanish language. This is the main reason both languages had to be separated from each other, and why sentences including both languages were removed from the dataset.

In this thesis, Phonemizer was applied to three datasets: Speech Commands, CSS10 and AOLME-Transcripts. Given that each dataset had their unique language and vocabulary, different sets of phonemes resulted. For Speech Commands, for example, the resulting phonemes (labels) were: `aɪ', `aʊ', `b', `d', `eɪ', `f', `h', `i', `iə', `iː', `j', `k', `l', `m', `n', `oʊ', `oːɹ', `p', `s', `t', `uː', `v', `w', `z', `æ', `ɑː', `ɑːɹ', `ɔ', `ə', `əl', `ɚ', `ɛ', `ɜː', `ɡ', `ɪ', `ɹ', `ʃ', `ʌ', `ʒ', `θ' and \emph{blank}, where \emph{blank} represents the default CTC's blank label.

It is important to mention that this is not a full list of English-based phonemes. Since Sphttps://www.overleaf.com/project/60ce5451dfbedc4d87d9b178eech Commands is only composed of 35 words, not all English phonemes are captured. For example, the phonemes `ð' and `ŋ' are not present in the aforementioned list, but pertain to the English words ``b ɹ ʌ ð ɚ'' (brother) and ``b ɪ l d ɪ ŋ'' (building), respectively.

On the other hand, for the Spanish dataset of AOLME-Transcripts, the list of labels was the following: `a', `aɪ', `aʊ', `b', `d', `e', `eɪ', `eʊ', `f', `i', `j', `k', `l', `m', `n', `o', `oɪ', `p', `pː', `r', `s', `t', `tʃ', `u', `w', `x', `ð', `ŋ', `ɔ', `ɛ', `ɡ', `ɣ', `ɲ', `ɾ', `ʝ', `β', \emph{blank} and \emph{white space}, where the \emph{white space} label represents the space between words.

Lastly, since CSS10 is based on Spanish from Spain, a different language was used when phonemizing. Therefore, in addition to the resulting labels of AOLME-Transcripts, two more were added to the mix: `ʎ' and `θ', where `ʎ' comes from words like ``b a t a ʎ a'' (batalla) and ``r o ð i ʎ a'' (rodilla); and `θ' comes from words like ``ɡ ɾ a θ j a s'' (gracias) and ``f w ɛ ɾ θ a'' (fuerza). In the order in which they were written, the respective translations to English are: battle, knee, thank you, and strength.

We would like to add that typing phoneme characters in LaTeX was not an easy task. To help others in the future, we provide the packages and commands used to achieve so (refer to Appendix \ref{appendices:phonemes}). Although it might not be an optimal solution, we hope it brings ease of pain to others.

\section{Learning Rate Scheduler} \label{sec:LRS}
The learning rate scheduler used in this thesis consists of a ``limited linear decay''. The idea was obtained from \cite{DLB} and consists of diminishing the learning rate until $\tau$ epochs have passed. The equation used to update the learning rate after epoch \emph{k} was the following: \begin{equation} e_{k} = (1 - \alpha)e_{0} + \alpha e_{\tau} \end{equation}.

\begin{figure}[h!]
 \centering
 \includegraphics[width=0.6\columnwidth]{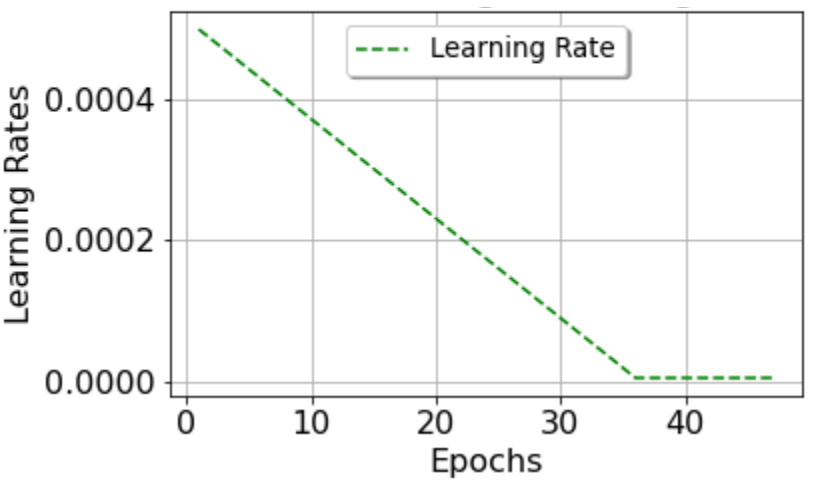}
    \caption[Learning Rate Progress]{Progress of learning rate using the ``limited linear decay'' scheduler.}
    \label{fig:LRprogress}
\end{figure}

Here $\alpha = k/\tau$, $e_0$ represents the initial learning rate, and $e_{\tau}$ the final learning rate. In all of our runs, we calculated $e_{\tau}$ by multiplying $e_0$ by 0.01. The variable $k$ represents the current epoch. Therefore, $e_k$ represents the learning rate at epoch $k$. Lastly, the variable $\tau$ represents the number of epochs to pass before the learning rate remains constant (at a value of $e_{\tau}$). Figure \ref{fig:LRprogress} shows the progress of a learning rate using this scheduler. In this example, $e_0$ was set to 0.0005 and $\tau$ to 35. Notice that once 35 epochs have passed, the learning rate remains constant.

\section{Baseline Keyword Classification Model} \label{sssec:dsArch}
In 2014, DeepSpeech \cite{ds} claimed to perform better than other systems in noisy environments (including commercial systems). Therefore, we decided to perform a study and determine whether or not DeepSpeech was able to recognize AOLME's noisy recordings. The study consisted of comparing DeepSpeech's performance against a simple SVM-PCA Classifier. The details and results of such an experiment are discussed below.

The first step was to develop a concise dataset suitable for the study. Its development is discussed in section \ref{sec:aolmeNums}. In brief, it was composed of 1,500 utterances from Speech Commands, 30 from TIMIT, and 36 from AOLME. The next step was to install and configure DeepSpeech; for which we used the 0.5.1 release. Transfer learning was then performed on such using the 1,500 utterances from Speech Commands. After this, the resulting model was used to run inferences on AOLME and TIMIT.

\begin{figure}[h!]
 \centering
 \includegraphics[width=0.8\columnwidth]{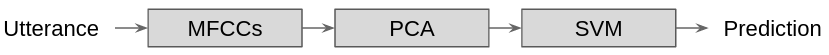}
    \caption[Diagram of SVM-PCA Classifier]{Diagram of SVM-PCA system. This system was used to compare performance against DeepSpeech on AOLME.}
    \label{fig:svmpca}
\end{figure}

In regards to the SVM-PCA Classifier, one can see its structure in Figure \ref{fig:svmpca}. Notice that first thirteen Mel-Frequency Cepstrum Coefficients (MFFCs) are calculated from each utterance. Where only twelve are kept (from the second to the thirteenth). Then, Principal Component Analysis (PCA) is performed on these coefficients to diminish their dimensionality, but still keep most of their information.

Next, the data is run through a OneVsRestClassifier using an SVC estimator (the SVM block in the diagram). To optimize the hyper-parameters of the system, a nested cross-validation procedure was utilized. During training, the system would output a score for each fold. During testing on the other hand, with the \emph{predict} method, the system would output labels' predictions of each tested utterance. The overall results of this experiment are shown in section \ref{results:baseline}.

\chapter{Results} \label{sec:results}
This section presents the results obtained by our low complexity network. We also discuss the parameters used for each model, including early stopping techniques. The datasets used to train and test the network were Speech Commands, CSS10, AOLME-Transcripts, and AOLME-Sentences. At the end of this section, we provide a comparison against commercial systems on AOLME-Sentences.

\section{Baseline with DeepSpeech and SVM-PCA Classifier} \label{results:baseline}
Results obtained from DeepSpeech and SVC-PCA are provided in Table \ref{tab:dsVSsvm}. Accuracy for DeepSpeech was calculated using $Acc = 100 * (1 - WER)$, where WER ranged from 0 to 1 given the similarity of the prediction to the target (0 being the same, 1 being complete opposites). On the other hand, accuracy for the SVM-PCA system was calculated using $Acc = \frac{Correct Predictions} {Number Of Samples}$. One can see in the table that, even with a stricter criterion, the SVM-PCA performed better than DeepSpeech when dealing with noisy recordings. In the case that this was not the outcome, we considered replacing PCA with Independent Component Analysis (ICA) since it is known to successfully remove statistical dependence between features \cite{balu}. Nonetheless, the results of Table \ref{tab:dsVSsvm} demonstrated the need to perform further research of speech recognition in collaborative learning environments.

\begin{table}[h!]
\caption[DeepSpeech vs. SVM Classifier]{Speech recognition accuracy obtained from DeepSpeech and SVM-PCA Classifier on TIMIT and AOLME.}
\label{tab:dsVSsvm}
\centering
 \begin{tabular}{| c | c | c| }
 \hline
 \textbf{Dataset} & \textbf{DeepSpeech} & \textbf{SVM-PCA Classifier} \\
 \hline\hline
 TIMIT & 73\% & 53\% \\
 AOLME-Numbers & 5\% & 36\% \\
 \hline
 \end{tabular}
\end{table}

\section{Performance on Speech Commands} \label{results:SC}
The first performance assessment of our low-complexity model was in the English language, using the Speech Commands v2 dataset. The test was performed on the 35-word task. Therefore, the data was split in accordance to the files \emph{validation\_list.txt} and \emph{testing\_list.txt}, as described in section \ref{datasets:sc}. In the end, the training, validation, and testing datasets were composed approximately of 24, 2.5, and 3 hours, respectively.

As discussed in section \ref{methodology}, the main structure of the model consisted of CNNs and GRUs. Using the python code shown in Appendix \ref{appendices:network}, for this assessment, the specifications of the network were the following. Regarding the CNN, four filters were used with 3-by-3 kernels and a stride of 2-by-2. Regarding the GRU, two bidirectional layers were used with 64 input units and 64 hidden units. In the end, the network used 155,353 parameters.

\begin{figure}[h!]
 \centering
 \includegraphics[width=0.7\columnwidth]{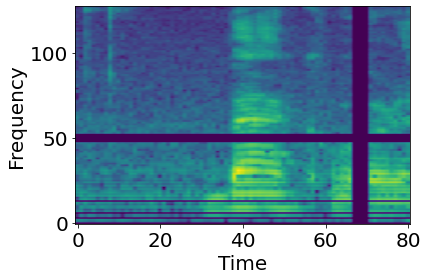}
    \caption[Example of SpecAugment Implementation]{Masked spectrogram of the word `backward' from Speech Commands using SpecAugment.}
    \label{fig:SC_specaugment}
\end{figure}

We used 41 labels (phonemes), as discussed in section \ref{sec:phonemizer}. We also used the learning rate scheduler described in section \ref{sec:LRS}, where $e_0$ was set to 5e-4 and $\tau$ to 40. Additionally, SpecAugment's frequency and time masking techniques were implemented. In Figure \ref{fig:SC_specaugment} we show an example of a masked spectrogram. It pertains to the word ``backward'' of the Speech Commands dataset.

The mask in the frequency domain (rows) was given by the boundary `FM'. In this particular example, the length of the frequency domain was 128 and FM was set to 14. Therefore, no more than 14 rows could have been masked in the spectrogram. On the other hand, a variable `TM' was used to assign a boundary to the time domain. A decimal value was assigned to this variable which was then multiplied by the length of the time domain. In this example, the length of the time domain was 81 and the value of TM was set to 0.0625. Therefore, the maximum number of columns to be masked was 5 (0.0625 x 81).

In terms of performance, two different metrics were considered, PER and WER. Early stopping was applied by checking the PER. If it did not decrease (improve) by a factor of 1/1000 in six consecutive epochs, training was stopped. From now on, we will refer to this early stopping technique as the \emph{common technique}.

\begin{figure}[h!]
 \centering
 \includegraphics[width=0.7\columnwidth]{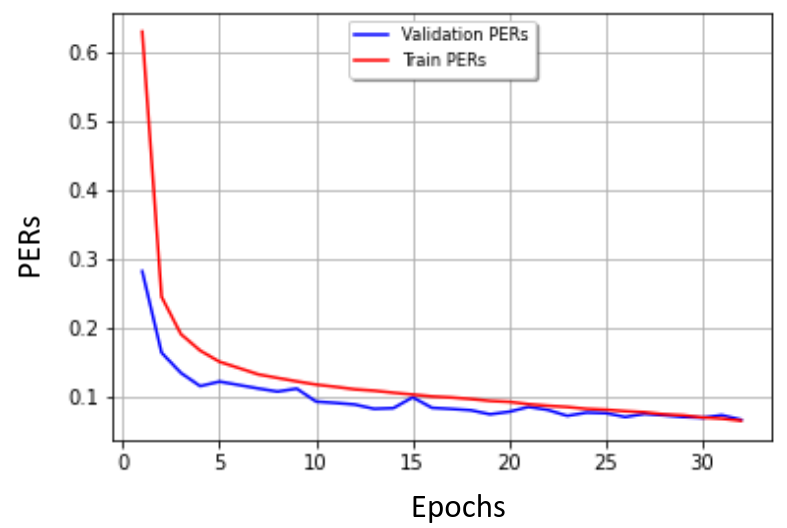}
    \caption[Training PER vs. Validation PER on Speech Commands]{Training and validation PERs per epoch on Speech Commands.}
    \label{fig:SC_PERs}
\end{figure}

Using the GPU discussed in section \ref{sec:netArch}, this model ran for 4 hours and 15 minutes before early stopping in epoch 32. For the validation set (\emph{validation\_list.txt}), it obtained a PER and WER of 0.0695 and 0.0864, respectively. For the testing set (\emph{testing\_list.txt}), 0.0801 and 0.0990 were obtained for PER and WER, respectively. In Figure \ref{fig:SC_PERs}, we show the process of the validation and training PERs per epoch. 

In Table \ref{tab:scSOTA}, we compare our results with those of SOTA systems on the 35-word task of Speech Commands v2. In their works, \cite{res15} and \cite{kwt3} used each word as a label, so accuracy is determined by the correct predictions divided by the total number of targets. In our work, accuracy is determined by subtracting the Word Error Rate from 1 and multiplying the result by 100, to provide the percentage of correctly classified words. We also show the number of labels and the number of parameters used by each system.

\begin{table}[h!]
\caption[Results on Speech Commands]{Comparison of results between SOTA systems and ours on the 35-word task of Speech Commands v2.}
\label{tab:scSOTA}
\centering
 \begin{tabular}{| c | c | c | c |}
 \hline
 \textbf{System} & \textbf{Accuracy} & \textbf{Num. of Parameters} & \textbf{Num. of Labels}\\
 \hline\hline
 kwt3 \cite{kwt3} & 97.69\% & 5.4M & 35\\
 res15 \cite{res15} & 96.40\% & 109k & 35\\
 ours & 90.10\% & 155k & 41\\
 \hline
 \end{tabular}
\end{table}

Although our accuracy was not as high as theirs, it shows that our approach has potential and is worth further analyzing. In particular, we use a similar amount of parameters as res15 (and much less than kwt3) and more labels than both of them.

\section{Training with CSS10} \label{sec:resultsCSS10}
The second assessment performed on our low-complexity model was on the Spanish language. The model was trained on the 2,308 simulated recordings of CSS10 (approximately equivalent to 5 hrs.), discussed in section \ref{datasets:css10}. The model was tested on the Spanish version of AOLME-Sentences and it targeted 40 different labels (including `ʎ' and `θ' as defined in section \ref{sec:phonemizer}).

With an amount of 88,828 parameters, the best results were obtained using one 2D CNN (with 16 filters, 3x3 kernels, and a stride of 2x2) and three bidirectional GRU layers (with 32 input units and 32 hidden units). The initial learning rate ($e_0$) was set to 4e-4 and $\tau$ was set to 30. In terms of data augmentation, SpecAugment was not used, since recordings had been simulated and augmented with Pyroomacoustics.

\begin{figure}[h!]
 \centering
 \includegraphics[width=0.7\columnwidth]{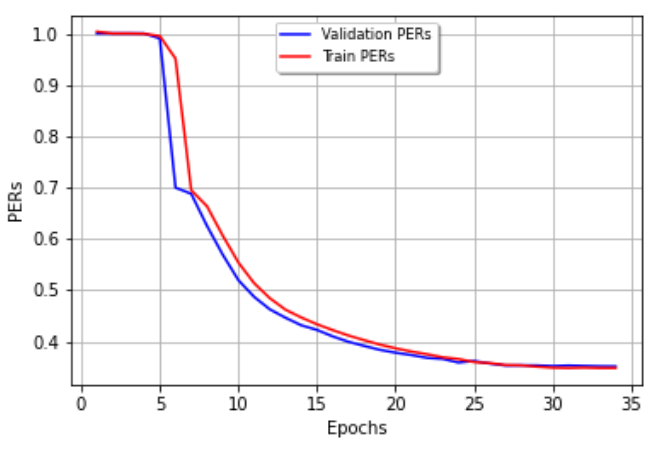}
    \caption[Training PER vs. Validation PER on CSS10]{Training and validation PERs per epoch on the CSS10 dataset.}
    \label{fig:CSS10_PERs}
\end{figure}

Early stopping was applied by keeping track of the validation PER, with the \emph{common technique}. If such did not decrease (improve) by a factor of 1/1000 in four consecutive epochs, training was stopped. Using the same GPU as before, the model ran into early stopping after 34 epochs (2 hours and 20 minutes). Figure \ref{fig:CSS10_PERs} depicts the validation and training PERs during the 34 epochs. This model achieved a 0.8463 PER on AOLME-Sentences which is equivalent to a 15.37\% phoneme accuracy.

\section{Training and Testing with AOLME} \label{results:AOtranscr}
The third and last assessment was performed using the Spanish recordings of AOLME-Transcripts (section \ref{sec:aolmeTranscr} describes this dataset in detail). We discuss the exhaustive number of runs performed by the model, present results obtained on AOLME-Sentences, and compare performance against commercial systems.

\subsection{Observations on a New Metric, \emph{Ratio PER}}
At first, we hypothesized that training jointly on CSS10 and AOLME Transcripts would produce better results. However, after multiple runs were performed and transfer learning combinations were implemented, it was clear that CSS10 was not improving performance. It was affecting it instead. We claim that this happened due to the Spaniard accent and the two additional phonemes. Therefore, hereafter, we continued to train the model purely on AOLME-Transcripts.

In these new runs, we observed that the checkpoint that was being saved using solely the validation PER was not providing the best generalization results. Therefore, we decided to implement a criterion that would save a second checkpoint. Such a criterion is described below.

\begin{figure}[h!]
 \centering
 \includegraphics[width=0.7\columnwidth]{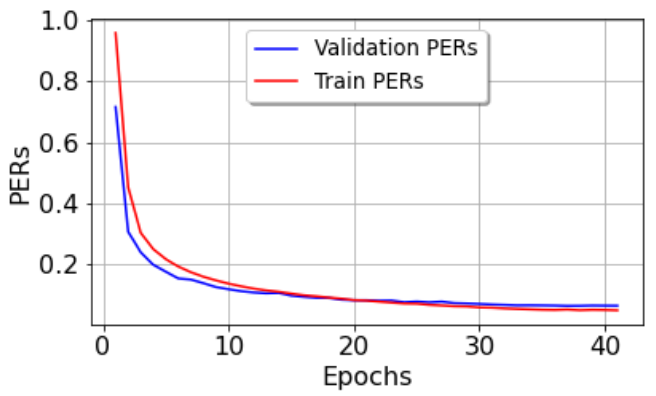}
    \caption[Training Progress on AOLME-Transcripts]{Plot of validation PER vs. training PER per epoch. Model was trained on AOLME-Transcripts.}
    \label{fig:ratioPER}
\end{figure}

The early stopping \emph{common technique} that we used would save a checkpoint once the metric did not improve by a certain percentage after some epochs. Although functional, we noticed that in some runs the validation PER would continue to improve even though ``convergence'' was already achieved. We define convergence as the point at which the validation and training metrics meet. We provide an example of convergence in Figure \ref{fig:ratioPER} where the model was trained on the Spanish version of AOLME-Transcripts. It shows the validation and training PERs per epoch.

In this example, early stopping occurred on epoch 41. Notice that between epochs 17 and 23, the training PER transitioned from being on top, to being below the validation PER. This transition period is a perfect example of convergence. We hypothesized that if we were to save a checkpoint during convergence, a better generalization score would be obtained and run time would be saved.

To prove this theory we used a third metric, namely \emph{ratio PER}. It was calculated by dividing  the training PER by the validation PER on each epoch. If the result was higher than 1, it meant that the training metric was above the validation metric. Conversely, if the result was lower than 1, the training metric was below the validation metric. Therefore, in order to ensure we saved a checkpoint during convergence, we used two threshold values for the \emph{ratio PER}, 0.98 and 1.02. For as long as the model's \emph{ratio PER} was between these two values, the model kept track of the validation PERs pertaining to those epochs. Once the model exited the convergence zone, a checkpoint pertaining to the lowest validation PER was saved.

As an example, we reference Figure \ref{fig:ratioPER} once again. Here, the only two epochs that reached convergence were epochs 18 and 19. The ratio PER of epoch 18 was 1.0007 and the ratio PER of epoch 19 was 0.9990. The validation PERs were 0.0919 and 0.0861, respectively. Since 0.0861 represented a better PER, a checkpoint was saved using the weights of epoch 19. Additionally, a second checkpoint was saved at epoch 37 (using the early stopping \emph{common technique}).

Both checkpoints were then assessed using the Spanish version of AOLME Sentences. The checkpoint from epoch 19 outputted a 0.7677 PER while the epoch from epoch 37 outputted a PER of 0.7684. After this, we decided to perform more comparisons and observed that with \emph{ratio PER} one can increase performance by 0.05-0.20\%. We also noticed that if we used \emph{ratio PER} to perform early stopping, run time could decrease considerably. In the aforementioned example, run-time would have decreased by a factor of 2. The model would have run 19 epochs instead of 37. Therefore, from here on, we performed early stopping with \emph{Ratio PER}. 

\subsection{Training and Testing}\label{sec:TT}
For the last set of runs, we used and compared two different augmentation techniques, Pyroomacoustics and SpecAugment. For Pyroomacoustics, we used 28,000 out of 37,272 utterances. For SpecAugment, we used the original 9,318 utterances, since SpecAugment changed masking locations after each epoch. In each run, AOLME-Transcripts was split in three, 80\% for training and 10\% for validation and testing.

The system configuration was the following, one 2D CNN (with 8 filters, 3x3 kernels, and a stride of 2x2) connected to two bidirectional GRU layers (using 64 input units and 64 hidden units). Number of labels was 36 (as discussed in section \ref{tab:phonemizerWords}), initial learning rate was 5e-4 and the $\tau$ coefficient was set to 35. This resulted in a total of 171,022 parameters.

Results from three different runs are shown in Table  \ref{tab:specVSpyroom}. TM and FM represent the time and frequency masking coefficients for SpecAugment. Validation and Test PERs represent the Phoneme Error Rates obtained on the test set of  AOLME-Transcripts. The last column shows the results obtained on the Spanish recordings of AOLME-Sentences.

\begin{table}[h!]
\caption[Comparison Between SpecAugment and Pyroomacoustics]{Comparison of results after training our model on different augmentation techniques. Notice the different values of TM and FM for SpecAugment. Keep in mind that the lower the PER, the better.}
\label{tab:specVSpyroom}
\centering
 \begin{tabular}{| c | c | c | c | c | c |}
 \hline
 \textbf{Augmentation} & \multirow{2}{*}{\textbf{FM}} & \multirow{2}{*}{\textbf{TM}} & \textbf{Validation} & \textbf{Test} & \textbf{PER on}\\
 \textbf{Technique} &  &  & \textbf{PER} & \textbf{PER} & \textbf{AOLME-Sentences}\\
 \hline\hline
 SpecAugment & 14 & 0.0625 & 0.0682 & 0.0696 & 0.7695\\
 SpecAugment & 14 & 0.125 & 0.0861 & 0.0882 & 0.7677\\
 Pyroomacoustics & - & - & 0.0647 & 0.0654 & 0.7208\\
 \hline
 \end{tabular}
\end{table}

One can see that Pyroomacoustics performed better than SpecAugment when validating and testing the model on AOLME-Transcripts. Additionally, Pyroomacoustics also showed better performance when testing on AOLME-Sentences (recordings directly extracted from AOLME students' interactions). This is important to us because it shows that AOLME's room simulations decrease the gap between synthesized and real audios by 17\%, in terms of Phoneme Error Rate.

\subsection{Comparison Against Commercial Systems} \label{sec:commSystems}
We decided to compare our results with the speech-to-text service of different commercial systems. In this research, we consider Microsoft, Google, IBM, and Amazon which systems were discussed in section \ref{sec:BGcomsyst}. Each of these systems was assessed on the Spanish version of AOLME-Sentences. The predicted transcripts were compared against the original counterpart and character error rates (CERs) were calculated (given that their systems outputted characters).

Since some predictions included uppercase letters and special characters, we replaced the first ones with lower cases letters and removed the special characters. We performed this step so that their predictions were aligned with the original transcripts and their performance was not affected. In Table \ref{tab:e2eSOTA}, we show the results obtained from each system. We also include the results obtained from our Pyroomacoustics model (mentioned in section \ref{sec:TT}). The accuracy presented in the table was calculated by subtracting the CER from 1 and multiplying by 100 (PER was subtracted from 1 in our case).

\begin{table}[h!]
\caption[Comparison Between Commercial Systems and Our Work]{Comparison of results between our work and the Speech-to-text service of different commercial systems on the 516 utterances of the Spanish version of AOLME-Sentences. Label accuracy is calculated by subtracting the label error rate from one and multiplying by 100. In the last column we specify whether the system used a language model or not. Additionally, notice that the numbers in parenthesis represent the comparison in multiples of the number of parameters and dataset size that we used.}
\label{tab:commercial}
\centering
 \begin{tabular}{| l | l | l | l | l |}
 \hline
 \multirow{2}{*}{\textbf{System}} & \textbf{Label} & \textbf{Num. of} & \multirow{2}{*}{\textbf{Dataset Size (hrs.)}} & \multirow{2}{*}{\textbf{LM}} \\
  & \textbf{Acc.} & \textbf{Parameters} & & \\
 \hline\hline
 \multirow{4}{*}{Google} & \multirow{4}{*}{27.28\%} & \multirow{2}{*}{2.5M \cite{streamingrnnt} ($15\times$)} & 4,000 \cite{streamingrnnt} ($300\times$) & \multirow{4}{*}{No} \\
  & & & 27,500 \cite{mobilernnt} ($2,110\times$) & \\
  & & \multirow{2}{*}{117M \cite{mobilernnt} ($680\times$)} & 12,500 \cite{las2} ($960\times$) & \\
   & & & 300 \cite{specaugment} ($20\times$) & \\
 \hline
 1CNN-2GRU & \multirow{2}{*}{27.92\%} & \multirow{2}{*}{171k ($1\times$)} & \multirow{2}{*}{13 ($1\times$)} & \multirow{2}{*}{No} \\
 + Pyroom & & & & \\
 \hline
 \multirow{2}{*}{IBM} & \multirow{2}{*}{31.80\%} & 28.5M \cite{ibmctc} ($165\times$) & 300 \cite{ibmlstm1} \cite{ibmlstm2} \cite{ibmctc} ($20\times$) & \multirow{2}{*}{No} \\
  & & 280M \cite{ibmctc} ($1,635\times$) & 960 \cite{ibmlstm1} \cite{ibmlstm2} \cite{ibmctc} ($70\times$) & \\
 \hline
 \multirow{3}{*}{Amazon} & \multirow{3}{*}{56.06\%} & \multirow{3}{*}{-} & 23,000\cite{minimumrnnt} ($1,765\times$) & \multirow{3}{*}{Yes} \\
  & & & 960 \cite{subwordrnnt} ($70\times$) & \\
  & & & 20,000 \cite{subwordrnnt} ($1,535\times$) & \\
 \hline
 \multirow{2}{*}{Microsoft} & \multirow{2}{*}{78.73\%} & 38M \cite{azure} ($220\times$) & \multirow{2}{*}{2,000 \cite{azure} ($150\times$)} & \multirow{2}{*}{Yes}\\
  & & 65M \cite{azure} ($380\times$) & & \\
 \hline
 \end{tabular}
\end{table}

Notice that Amazon's number of parameters is blank since we could not gather a concrete number. However, we expect their models to train on the millions since their network architectures like \cite{minimumrnnt} and \cite{subwordrnnt}, used a total of seven LSTM layers with 1,024 hidden units each.

In comparison to Google's Speech-to-text service (whether it is based on an RNN-T or a LAS architecture), we claim that when targeting collaborative learning environments, our approach is competitive. Not only is a better accuracy obtained, but the low-complexity aspect of our network reduces the number of parameters, utterances and GPUs needed. For instance, our network uses less than 180k parameters while theirs use more than 2.5M. We use less than 13 hrs. for training while they use 300 or more. In terms of GPUs, systems like SpecAugment \cite{specaugment} were trained on 32 Google Cloud TPU chips for seven days, while we trained our model in a single 8GB GeForce GTX 1050 Mobile for three hrs. and 45 minutes.

In regards to IBM's speech-to-text service, we observe that a slightly higher accuracy was obtained in comparison to us. We claim, however, that we can improve our results by properly optimizing the network and simulation framework (as discussed in chapter \ref{sec:conclusions}).

Due to the large gap between IBM's and Google's results from Amazon's, we argue that a language model or post-processing technique must be taking place in their service and works well for small vocabularies like the one of AOLME-Sentences (composed only of 368 different words). Therefore, we claim that our model is not directly comparable to theirs since we have not implemented language models or post-processing techniques.

Similarly, we are certain that the system behind Microsoft implements LMs (like LSTM-LM and ngram-LM) and extensive post-processing techniques (like LSTM rescoring, ngram rescoring, and backchannel penalty) as it is described in \cite{azure}. Therefore, we cannot be compared to their system as well since we have not implemented any of these techniques.

\chapter{Conclusions and Future Work} \label{sec:conclusions}
In this work, we have presented a phoneme-based low-complexity network capable of training in one or two languages. Using less than 180k parameters and 13 hrs. of data, it can compete against speech-to-text systems like Google's and IBM's. We have also described a simulation and augmentation framework. It is capable of mimicking a noisy classroom and expanding a dataset up to a factor of four. Additionally, we show that it can outperform SOTA data augmentation techniques (like SpecAugment) in collaborative learning environments.

Furthermore, we have introduced an audio dataset obtained from live interactions of middle school students in classrooms. We also discuss the obstacles we faced when gathering this dataset, particularly the ones related to choosing the starting and ending boundaries. Although we present two different programs that facilitate this procedure, we reiterate that this task remains tedious and time-consuming.

For future work, we plan to finalize the English version of AOLME-Transcripts by thoroughly checking the produced recordings from TTSMP3. Then, after running the recordings through our simulation and augmentation framework, we will train our model on such and assess performance on the English version of AOLME-Sentences. Consequently, this will allow us to train and test our model in both languages at the same time.

Additionally, we intend to optimize our model by implementing cross-validation and studying language models and post-processing techniques. We also plan to optimize our simulation framework by determining more accurate speaker geometries.

Lastly, we intend to use autoencoders for self-supervised learning. Given that most of AOLME's recordings are unlabeled, this approach will impact our work since there will not be a need for text transcriptions. Therefore, once the autoencoder learns from the unlabeled data, we can use the resulting weights to train on the limited amount of labeled data.

\chapter*{Appendices}

\addcontentsline{toc}{chapter}{Appendices}
   \contentsline {chapter}{\numberline {A}Python code to automatically download audios}{56}
   \contentsline {chapter}{\numberline {B}Python code to format audios using ffmpeg}{58}
   \contentsline {chapter}{\numberline {C}Network architecture of the proposed model in Python}{59}
   \contentsline {chapter}{\numberline {D}Python sample on how to train the network}{63}
   \contentsline {chapter}{\numberline {E}Python code to phonemize text}{65}
   \contentsline {chapter}{\numberline {F}How to Type IPA Phonemes in LaTeX}{66}

\appendix
\chapter{Python code to automatically download audios} \label{appendices:selenium}
    {
\begin{lstlisting}[language=Python]
import os
from selenium import webdriver
from selenium.webdriver.support.ui import Select
import time

#Declare Variables
Login_Xpath = '/html/body/div[2]/form/input[3]'

#Grab Firefox's driver, open browser and access link
driver = webdriver.Firefox()
driver.get('https://ttsmp3.com/login.php')
#Input username and password in respective fields
username = driver.find_element_by_name('username')
username.send_keys('my_username')
password = driver.find_element_by_name('my_password')
password.send_keys('my_password')
#Click login button
login_button = driver.find_element_by_xpath(Login_Xpath)

#Locate 'language/voice' dropdown menu and choose 'Joey'
select = Select(driver.find_element_by_id("sprachwahl"))
select.select_by_value('Joey')

time.sleep( 1.5 ) #Give it time to select field
#Locate text area and type "Hello World!"
text_area = driver.find_element_by_id("voicetext")
text_area.send_keys("Hello World!")

#Locate and click download button
download_button = \
    driver.find_element_by_id("downloadenbutton")
download_button.click()	
time.sleep( 1.5 ) #Give it time to download
#Clear text area
text_area.clear()
#Close browser
driver.close() 
\end{lstlisting}}

\chapter{Python code to format audios using ffmpeg} \label{appendices:ffmpeg}
    {
\begin{lstlisting}[language=Python]
import subprocess as subp   
#Specify source and destination paths
audio_input_path = "..."
audio_output_path = "..."
#Remove ffmpeg verbose
cmd = "ffmpeg -hide_banner -loglevel error -i "
#Path to audio that will be converted
cmd += f"'{audio_input_path}' "
#Specify codec to use for audio
cmd += "-acodec pcm_s16le "
#Specify number of channels and sample rate
cmd += "-ac 1 -ar 16000 "
#Path to save resulting audio
cmd += f"{audio_output_path}"
#Run command
subp.run(cmd, shell=True)
\end{lstlisting}} 
    
\chapter{Network architecture of the proposed model in Python} \label{appendices:network}
    {
\begin{lstlisting}[language=Python]
'''/***************************************************
    File: models.py
    Author(s): Mario Esparza, Luis Sanchez
    Date: 02/26/2021
    
    Network architecture used to train the model.
    Based on this implementation: Building an 
    end-to-end Speech Recognition model in PyTorch
    by Michael Nguyen.
****************************************************'''
import torch.nn as nn
import torch.nn.functional as F

class BiGRU(nn.Module):

    def __init__(self, gru_dim, gru_hid_dim, \
        gru_layers, dropout, batch_first):
        super(BiGRU, self).__init__()

        self.BiGRU = nn.GRU(
            input_size=gru_dim, 
            hidden_size=gru_hid_dim,
            num_layers=gru_layers, 
            batch_first=batch_first, 
            bidirectional=True
        )
        
        self.layer_norm = nn.LayerNorm(gru_dim)
        self.dropout = nn.Dropout(dropout)

    def forward(self, x):
        x = self.layer_norm(x)
        x = F.gelu(x)
        x, _ = self.BiGRU(x)
        x = self.dropout(x)
        return x

class SpeechRecognitionModel(nn.Module):
    def __init__(self, hparams):        
        super(SpeechRecognitionModel, self).__init__()
        self.cnn1_kernel = hparams['cnn1_kernel']
        self.cnn1_stride = hparams['cnn1_stride']
        
        # cnn for extracting heirachal features
        self.cnn = nn.Conv2d(
            1, 
            hparams['cnn1_filters'], 
            hparams['cnn1_kernel'], 
            stride=hparams['cnn1_stride']
        )
                               
        in_feats = hparams['cnn1_filters'] * \
            (hparams['n_mels'] // hparams['cnn1_stride'])
        self.fully_connected = nn.Linear(in_feats, \
            hparams['gru_dim'])
        
        self.birnn_layers = BiGRU(
            hparams['gru_dim'], 
            hparams['gru_hid_dim'],
            hparams['gru_layers'], 
            hparams['gru_dropout'], 
            batch_first=True
        )

        #Set up value for classifier's out-ftrs of fc layers
        fc_out_ftrs = 0
        
        if hparams['gru_hid_dim']*2 <= hparams['n_class']:
            fc_out_ftrs = hparams['n_class']
        else:
            fc_out_ftrs = (hparams['gru_hid_dim']*2 \
                - hparams['n_class']) // 2
            fc_out_ftrs += hparams['n_class']
        
        self.classifier = nn.Sequential(
            nn.Linear(hparams['gru_hid_dim']*2, fc_out_ftrs),
            nn.GELU(),
            nn.Dropout(hparams['dropout']),
            nn.Linear(fc_out_ftrs, hparams['n_class'])
        )

    def add_paddings(self, x):
        #Pad if necessary
        paddings = [0,0,0,0] #left, right, top, bottom
        sizes = x.size()
        if (sizes[2] % 2) != 0: #H
            paddings[3] += 1
            
        if (sizes[3] % 2) != 0: #W
            paddings[1] += 1
        
        #Equation below won't work if cnn1_stride != 2
        pad_val = int(self.cnn1_kernel/self.cnn1_stride-0.5)
        paddings = [val + pad_val for val in paddings]
        self.pad = nn.ZeroPad2d(paddings)

    def forward(self, x):
        x = self.pad(x)
        x = self.cnn(x)
        sizes = x.size()
        x = x.view(sizes[0], sizes[1] * sizes[2], sizes[3])
        x = x.transpose(1, 2) # (batch, time, feature)
        x = self.fully_connected(x)
        x = self.birnn_layers(x)
        x = self.classifier(x)
        return x
\end{lstlisting}}
    
\chapter{Python sample on how to train the network} \label{appendices:train}
    {
\begin{lstlisting}[language=Python]
import torch
import torch.nn as nn

import constants as cnstnt
from models import SpeechRecognitionModel
from utils import train, dev, LR_SCHED
    
#Initialize Hyper Parameters
hparams = {  
    'cnn1_filters': [8],
    'cnn1_kernel': [3],
    'cnn1_stride': [cnstnt.CNN_STRIDE],
    'gru_dim': [64],
    'gru_hid_dim': [64],
    'gru_layers': [2],
    'gru_dropout': [0.1],
    'n_class': [-1], #dynamically initialized later
    'n_mels': [128],
    'dropout': [0.1], #classifier's dropout
    'e_0': [5e-4], #initial learning rate
    #Set 'T' to -1 for steady LR throughout training
    'T': [35], 
    'bs': [2], #batch size
    'epochs': [60]
}
#Initialize model, optimizer and CTC loss function  
model = SpeechRecognitionModel(hparams)
optimizer = torch.optim.AdamW(
    model.parameters(),
    hparams['e_0']
)
criterion = nn.CTCLoss(blank=blank_label)
#Initialize custom learning rate scheduler
scheduler = LR_SCHED(hparams)
#Train and validate
for epoch in range(1, hparams['epochs'] + 1):
    train(model, device, train_loader, criterion,
    	optimizer, scheduler, epoch, ...)
    
    dev(model, device, dev_loader, criterion,
        epoch, ...)
            
\end{lstlisting}}
   
\chapter{Python code to phonemize text} \label{appendices:phonemizer}
    {
\begin{lstlisting}[language=Python]
from phonemizer import phonemize
from phonemizer.separator import Separator

#Custom separator, so phonemes are separated by a white space
sep = Separator(phone=' ', syllable='', word='_') 
#Example: Phonemizing in Spanish using IPA backend (espeak)
word = "baile"
L = 'es-419'
B_E = 'espeak'
phones = phonemize(word, L, B_E, sep)[:-2]
print(phones)
#Example: Phonemizing in English using IPA backend (espeak)
word = "barbecue"
L = 'en-us'
B_E = 'espeak'
phones = phonemize(word, L, B_E, sep)[:-2]
print(phones)
\end{lstlisting}}
   
\chapter{How to Type IPA Phonemes in LaTeX} \label{appendices:phonemes}
\begin{lstlisting}[language=TeX]
%Needed Packages:
  \usepackage{tipa}
  \usepackage[T1]{fontenc}
  
%Helpful Links:
  %http://detexify.kirelabs.org/classify.html
  %http://math.colorado.edu/~rohi1040/symbols-a4.pdf   %page9
  %https://en.wikipedia.org/wiki/Phonetic_symbols_in_Unicode
  
%From package tipa
  \DeclareUnicodeCharacter{029D}{\textctj} %ʝ
  \DeclareUnicodeCharacter{03B2}{\textbeta} %β
  \DeclareUnicodeCharacter{0283}{\textesh} %ʃ
  \DeclareUnicodeCharacter{025B}{\textepsilon} %ɛ
  \DeclareUnicodeCharacter{027E}{\textfishhookr} %ɾ
  \DeclareUnicodeCharacter{02D0}{\textlengthmark} %ː
  \DeclareUnicodeCharacter{0272}{\textltailn} %ɲ
  \DeclareUnicodeCharacter{0259}{\textschwa} %ə
  \DeclareUnicodeCharacter{026A}{\textsci} %ɪ
  \DeclareUnicodeCharacter{0251}{\textscripta} %ɑ
  \DeclareUnicodeCharacter{0279}{\textturnr} %ɹ
  \DeclareUnicodeCharacter{028A}{\textupsilon} %ʊ
  \DeclareUnicodeCharacter{0254}{\textopeno} %ɔ
  \DeclareUnicodeCharacter{0263}{\textgamma} %ɣ
  \DeclareUnicodeCharacter{0261}{\textscriptg} %ɡ
  \DeclareUnicodeCharacter{025A}{\textrhookschwa} %ɚ
  \DeclareUnicodeCharacter{025C}{\textrevepsilon} %ɜ
  \DeclareUnicodeCharacter{028C}{\textturnv} %ʌ
  \DeclareUnicodeCharacter{0292}{\textyogh} %ʒ
  \DeclareUnicodeCharacter{03B8}{\texttheta} %θ
  \DeclareUnicodeCharacter{028E}{\textturny} %ʎ
  
%From package [T1]{fontenc}
  \DeclareUnicodeCharacter{00E6}{\ae} %æ
  \DeclareUnicodeCharacter{00F0}{\dh} %ð
  \DeclareUnicodeCharacter{00A0}{\ng} %ŋ
\end{lstlisting}


\end{document}